\documentclass[english]{article}
\usepackage[T1]{fontenc}
\usepackage[latin9]{inputenc}
\usepackage{graphicx}

\makeatletter
\newcommand{\lyxaddress}[1]{
\par {\raggedright #1
\vspace{1.4em}
\noindent\par}
}

\makeatother

\usepackage{babel}
\begin{document}

\title{A non iterative method of separation of points by planes in n dimensions and its
application}

\author{K.Eswaran}

\maketitle

\lyxaddress{Dept. of Computer Science, Sreenidhi Institute of Science and Technology,
Yamnampet, Ghatkesar, Hyderabad, 500010 India }
\begin{abstract}
We demonstrate an algorithm that can separate any number of points
in n-dimensional space from one another by planes. Given a set G of $N_{f}$ points, along with their coordinates, in n-dimensional X-Space, the algorithm partitions all the $N_{f}$ points by using planes such that no two points are left un-separated
by some plane. The algorithm proceeds by transferring points from G to another set S in n-dimensional X-space, to the same coordinate position in S as it was in G. However initially S has very few points and very few planes, but the few planes are so chosen that all the points in S are already separated from each other by the few planes contained in it.  Each point is then chosen at random from G and then transferred to S one by one, matters are so arranged that if necessary new planes are drawn in S so that the points in S will always be separated. In order to do this each point from G is labeled as a member of S if it is separated from others in S, otherwise it is labeled as a ``special'' 
point which is not yet separated. The collection proceeds (at random)
as new points are added to S from G and each new point is either separate and becomes a member of S or labeled as ``special''. As soon as $n$ ``special'' points are collected
they are all separated by a single extra plane, which is then added to
the collection. The algorithm is made possible because a new concept
called Orientation Vector is used. This vector is a Hamming vector
and is associated with each point and has all the information necessary
to ascertain if two points are separate or not. The algorithm proceeds
till S contains all the planes which separate all the points. The
method is non iterative, it will always halt successfully and the
algorithm strictly follows Shannon's principle of making optimal use
of information as it advances stage by stage. It has the property
of restart, if new points are needed to be separated the algorithm
can continue from where it left off. At some later stage if the dimension
of the data (n to n+r) is increased the algorithm can still continue
from where it left off, and tackle the new data points which are of
a higher dimension, after some minor modifications. 

A proof is provided with a worked example. The computational Complexity
is of $O(n.N log(N)) + O(n^3 log(N))$, where $N$ is the given number
of points and \$n\$ is the dimension of space. In summary this paper
describes a non-iterative algorithm of separating a given set of points
in n-dimension by planes. Its application to data retrieval problems
in very large medical data bases is also given. Possible future applications are also identified.
\end{abstract}

\section{Introducing the Algorithm}
We assume that we are given a set G consisting of $N_{f}$ points, along with 
their coordinates, in n-dimensional X-Space. Our task is to partition all the $N_{f}$
points by using planes such that no two points are left un-separated
by some plane. Our method is such that each point in G, is looked
at mostly once and once only, then it is put in a another set S which
consists of other points and planes, S will contain all the coordinate
information of the points in it as well as the equations of the planes
which are in it. We start with a set S which has a few number of points
$N=N_{0}$ which were drawn from G, but positioned in S such that they have the same coordinates in S as they had in G, and also, S, has initially an adequate
number of planes $q=q_{0}$, so chosen that they separate all the
points $N=N_{0}$ presently in S. After which points are chosen randomly
from G; a point thus chosen can be immediately put in S (but at a position which has the same coordinate as the point had in G) if it will
be separated from all other points in S by the planes presently in
S, if this is not so then the point will be put temporarily in a set
T. Points from G are thus picked and put in (S or T), one by one,
at some time it may be necessary to add a new plane such that, after
the inclusion of this new plane (making $q=q+1$) some $n$ points
currently in T are separated from one another by one or other of the
$q+1$ planes currently present, and thus can be transfered to S.
All this is only possible because we have discovered a quick method
which reveals whether a particular point P in S is separated from
other points such as R in S or not. We are able to do this because
we associate an \textquotedbl{}Orientation Vector\textquotedbl{} $Ov(P)$,
with every point P in S, this Orientation Vector is a Hamming Vector
of dimension $q$ the current number of planes in S. The Orientation
Vector has the property that two points will not have a plane between
them if and only if $Ov(P)=Ov(R)$. That is $Ov(P)\ne Ov(R)$ automatically
implies that P and R are separated by at least one of the $q$ planes
in S. Since Orientation Vectors are Hamming vectors of dimension $q$,
the condition $Ov(P)\ne Ov(R)$ can be quickly verified, to ensure
that points P and R are separable in S. The above process of including
points, one by one, chosen randomly from G is done in such a way that
all the points in S always remain separated from one another by the
current number of planes $q$, in S. This condition is ensured by
following a procedure which we will now describe. However, as related
earlier, we require the services of another set T, and we will use
this set T to temporarily hold certain {}``candidate'' points and
line segments till the {}``candidate'' points are qualified to be
transfered to S. We now describe the procedure: Let us say that S
contains N points and q planes, because of the process followed so
far all the points $N$ are separable by the $q$ planes in S. We
now choose randomly a new point, say P, from set G, this point P is
then considered as a \textquotedbl{}candidate'' for being the $N+1$
point for inclusion in S, however for P to be actually included in
S we must calculate the Orientation Vector $Ov(P)$ and ensure that
$Ov(P)\ne Ov(R),\forall$ points $R\in S$. Since there are $N$ points
such as R, the preceding conditions on $Ov(P)$ involve $N$ Hamming
vector comparisions. If P satisfies all these conditions then this
implies that P is separable from all points currently existing in
S and can thus be safely be added to S so P is transferd to S and
then $N\rightarrow N+1$ and a new \textquotedbl{}candidate point''
point P is chosen from G. However if the test for $Ov(P)$ fails this
means there is some point $A\in S$ s.t. $Ov(P)=Ov(A)$ , this indicates
that P and A are not separated by $q$ and thus P cannot be put in
S. At this point we rename the point P as B for convenience and put
B in T and we also store the {}``pair'' (A,B) in T. we will treat
(A,B) as a line segment connecting two points A and B where $A\in S$
and $B\notin S$, we then keep a Counter which keeps a count of the
number of pairs in T. After this we randomly choose another point
P from G as the next possible {}``candidate'' for S and repeat the
procedure. It is important that at each stage of the algorithmic process,
the Orientation Vector of points in S are calculated and stored in
S along with the \textquotedbl{}candidate\textquotedbl{} points and
their Orientation Vectors in another set T. The algorithm keeps adding
points and calculating planes to be put in S step by step all the
time ensuring that S contains only points that are separated from
one another by the $q$ planes presently contained in S. Each point
when thus chosen either is separate from other points or if not, as
we have seen matters are so arranged, that it is identified {}``as
a special point'' which is not yet separated, such points such as
B, are placed in T along with its {}``pair'' and with the segment
(A,B).

Matters can always be so arranged that each point in T has one and
only one point in S which is its neighbor. (Two points are neighbors
if they are un-separated). It is obvious that a point such as B cannot
have two points say A and C as neighbors, with both A and C belonging to S,
 because this means $Ov(B)=Ov(A)$ and $Ov(B)=Ov(C)$ which implies $Ov(A)=Ov(C)$
which is impossible because A and C, if they both belong to S, cannot
have their Orientation Vectors equal to each other since S contains
points which are separated from one another by $q$ planes.

 The collection proceeds (at random) as new points are added to S or T depending on
whether the new point is either separate or labelled as {}``special''.
As soon as n {}``special'' points are collected in T,the process
of collecting points is temporarily halted. Now the set T is examined,
since the Counter has stopped at $Counter=n$, there will be in T,
precisely $n$ pairs of points $(A_{j},B_{j}),(A_{j}\in S$ and $B_{j}\notin S$.
\\
\emph{We now have arrived at a very crucial stage in the algorithm},
till now we have been merely adding points, now it is time to add
a new plane to S. We do so now, and we do it in such a manner that
the new plane will pass through all the midpoints $m_{j}$ of all
the $n$ segments $(A_{j},B_{j})$ that we have collected in T. If
we do this, we would have in a single stroke, not only separated all
the $n$ points $B_{j}$ from their respective \textquotedbl{}pairs\textquotedbl{}
the $A_{j}$, by this new plane, but also from one another. It can
be easily proved that all the $B_{j}'s$ will be separated from one
another if we do not permit two $B's$ to be the neighbor of the same
A. (This situation is rare in large n dimension space but is discussed later 
and shown that it can be handled). Note this requirement automatically means the
two $B's$ are separated because if $B_{i}$ is a neighbor of $A_{i}$
and if $B_{j}$ is not a neighbor of $A_{i}$ this means $Ov(B_{i})=Ov(A_{i})$
and $Ov(B_{j})\ne Ov(A_{i})$ implying $Ov(B_{i})\ne Ov(B_{j})$.
Thus if we do not permit two $B's$ to be neighbors of the same $A$,
then after the $q+1$ plane bisects all the $n$ segments then we
can include the $q+1$ plane in S, thus enabling all the $n$ points
$B_{j},(j=1,2,..n)$ to be included in S too. The separation of $n$
line segments by a single extra plane, can be done only because we
are in n-dimension space. The calculation of the coefficients of this
$(q+1)^{st}$ plane can be done by (say) Gaussian elimination and
involves $O(n^{3})$ multiplications. Then the new plane with its
coefficients are added to S. All the $n$ recently separated points
$B_{j}$ are added to S so $N\rightarrow N+n$ after this all the
$B's$ are removed from T along with the pairs (because they are all
separate - a reconfirmation is done). And importantly, after adding
this $q+1$ new plane to S the Orientation Vectors of all points in
S have to be modified wrt this new plane all the Orientation Vectors
are now of dimension $q+1$ -it only means adding a bit to each vector.
\\
This noniterative algorithm is possible because we use the concept
of an Orientation Vector {[} 3{]} which is a Hamming vector and has
the following properties: (i) every point is associated with its own
Orientation Vector, (ii) the Orientation vector for a point P in S
keeps track of how P is positioned with respect to all the planes
in S and (iii) the Orientation Vector is unique for each point in
S and (iv) the Orientation Vectors of two points say P and Q are unequal
if and only if they are separated by at least one plane in S; equality
implies that P and Q are not separated.

The method is a kind of pigeon hole principle, except that each point
is a pigeon and the pigeon holes (region between half spaces) are
added in installments (much more than one at a time) only when they
are needed, to accommodate new pigeons. At each stage of the algorithm
there is only one pigeon in its hole, and as the algorithm proceeds
the size of some pigeon holes may reduce as they are partitioned, and  many pigeon holes are created and new pigeons are added to occupy some of the empty holes and at the end of the algorithm all the pigeons are accommodated.

\section{Setting up the Tasks at hand and necessary definitions}

We give the necessary definitions and briefly prove the fundamental
algorithm by which planes which separate a given set of points could
be found. 

Our task is to determine the equations of the planes that can separate
the points in such a way that every point is separated from another
by at least one plane.

Assumption: It is assumed that we have specified a (defined) normal
direction, a point which lies on the positive side of the normal is
said to lie on the positive side of the plane, on the other hand if
the point lies on the other side it is said to lie on the negative
side.This direction is easily found if the equation of the plane is
known for example if 
\begin{equation}
1+\alpha_{1}x_{1}+\alpha_{2}x_{2}+...,+\alpha_{n}x_{n}=0
\end{equation}

is the equation to some $n$- dimensional plane then a point P whose
coordinates are $(p_{1,}p_{2},....,p_{n})$ will be said to be on
the positive side if $1+\alpha_{1}p_{1}+\alpha_{2}p_{2}+...,+\alpha_{n}p_{n}>0$
and in the negative side if

$1+\alpha_{1}p_{1}+\alpha_{2}p_{2}+...,+\alpha_{n}p_{n}<0$

Each component of the Orientation Vector of a point P in X space,is on the positive side or negative side of each plane.(The`Orientation Vector', is defined in the next section, below.)

\subsection{Definitions and Terminology: }

\textbf{Orientation Vector}: Suppose we have a point P in n-dimension
space \textbf{(also called X-space)} and suppose we are given 3-planes.
we define a Orientation Vector as a Hamming vector whose components
precisely specify on which side the point P lies with respect to the
3 planes. Example: if the point P lies on the positive side of plane
1, negative side of plane 2 and positive side of plane 3 , then we
define the Orientation Vector associated with P as $Ov(P)$ and define
$Ov(P)\equiv(1,-1,+1)$. Similarly a point Q which lies on the negative
side of plane 1, negative side of plane 2 and positive side of plane
3 will have an Orientation Vector $Ov(Q)\equiv(-1,-1,1)$ . Points,
whose Orientation Vector differ from another by at least one component
can be said to be separated by at least one plane.

\textbf{Q Space or Hamming space: }We will also call the space spanned
by the Orientation Vector as Hamming Space or Q space. Since each
point P in X-space has an Orientation (Hamming) Vector associated
with it, we can imagine that all the points in X space are mapped
to a point in Hamming space. Of course this mapping is many to one,
but the fact to notice is the following: Let P be some point in X-Space,
then all points, R, in X space which are not separated from P by planes
will all have the same Orientation Vector as P ie. $Ov(P)=Ov(R)$
and we describe this by saying: {}``Points P and R belong to the
same `quadrant''' this statement is certainly true for the {}``images''
of P and R in Q-Space, but we will loosely use this terminology for
the points in X-space and say P and R are in the same {}``quadrant'',
what we really mean is that P and R are points in X-Space and that
they are not separated by planes; Sometimes we will use the term \textbf{neighbors}
and say P and R are neighbors. The point to remember is that P and
R will be neighbors if and only if $Ov(P)=Ov(R)$. Therefore if one
wishes to find out if two points A and B are separated in X-Space
which contains planes, all one needs to do is to compare their Orientation
Vectors: $Ov(A)$ with $Ov(B)$.

\textbf{A Saturated Plane:} We consider a plane in $n$ dimension
space to be {}``saturated'' if it has already been constrained to
pass through $n$ points and hence cannot be adjusted to pass through
a new point, the coefficients of such a plane are completely determinable.
Eg. A plane in 3 dimension gets saturated if it is made to pass through
three points.

We will suppose we are given all the data containing $N_{f}$ points, and that all 
all their coordinates in $n$ dimension space are known to us. Our task is then to find the planes numbering $q_f$, that can separate all these points from one another and also to find all the coefficients of the equations which define each one of these planes.

\subsection{Input Output Requirements}

\textbf{Input to the Algorithm} We are given a set G containing a
total of $N_{f}$ points in $n$ dimensional space. That is we know
the coordinates of all these points. These points may also belong
to different classes, so they may be given a label or `color'. However,
since we are isolating all points from one another regardless of their
class, the labels do not matter here, but will be useful in certain
applications.

\textbf{Without, too much of a loss of generality we will assume that the coordinates of all the N points in G are rational numbers, i.e they are either integers or fractions.} We will see that this assumptions makes the coefficients of  all the $q$ planes that separate the $N$ in G, also rational numbers. 

\textbf{Output Items of the Algorithm:} The equations of the planes
which divide each point in such a way that every point is separated
from another. And an integer $q_{f}$ which is equal to the
total number of planes required. The equations of the $q^{th}$ plane will be of the form:
\begin{equation}
\tau +\alpha_{q1}x_{1}+\alpha_{q2}x_{2}+....+\alpha_{qn}x_{n}=0\quad \quad (q= 1,2,..,q_f)
\end{equation}
 where all the $\alpha_{qj}$ are coefficients of the plane all these coefficients  are determined by the algorithm and  $\tau$ is a known number nearly equal to unity. 

\textbf{ Storage Requirements for running the algorithm:} A set Set
S which will contain a list of $N$ points and $q$ planes. S will contain the Orientation Vectors of all the points in S and the identification numbers (labels) for the planes, along with the coefficients which define each plane. S contains an array V for storing Orientation Vectors; and `Counter': An integer number. 

In all stages of the algorithm
S, will contain only those points, (N in number) which are completely
separated from one another by the q planes contained in S. This condition
of separability of the N points in S is never violated even though
N and and q change as the algorithm progresses. When the Algorithm
ends S will contain the necessary output.

Another Set T which contains points and their Orientation Vectors
and an array M :{}``List of Midpoints'' which are the coordinates
of the mid points of certain line segments. At each stage of the algorithm
this set T, will contain pairs of points one not in S and the other
in S. We are sure that each such point has only one such neighbor (in S),
because the points in S are always separate from one another and hence
two points in the same quadrant cannot both belong to S. Each such
point stored in T will also store the coordinate of the point which
is the midpoint of the line joining it to its neighbor and the line
segment with its neighbor. The points in T are the candidates waiting
to be put in S but they first have to be separated from all the points
in S. We shall see that as soon as T has a collection of n pairs of
such points, then a new plane is drawn which separates all these points
so that they become eligible to be incorporated in S. (Later on, we will permit a given point in S to have two or even three neighbors in T, these too will be candidates to be put in S, but for a first reading of this algorithm, it is simpler to assume that a point in S can have at most one neighbor in T.) 

We also need another set D, this is the `Dust-Bin' set, it removes
all `accumulation' points when detected from set G, (as explained
later this is done so that the algorithm does not go into an infinite
loop; which may happen if G contains a sequence of points converging to an `accumulation' point or `limit' point - we assume that such points do not exist in G).

\medskip{}

\subsection{Steps of the Algorithm}

Step 1: Initially collect a small number of initial points numbering
$N_{0}$ and choose a set of $q_{0}$ planes that separate each one
of these $N_{0}$ points and put these planes in S. The coefficients
defining the equations of these $q_{0}$ planes should be determined
(or randomly chosen but ensuring that they separate the $N_{0}$),
call $N=N_{0}$ and $q=q_{0}$. Store all Orientation Vectors of points
in S in array V. In addition we need a set T which will contain points
which cannot become immediate members of S but are prospective members
and will become members of S eventually. Set T will contain points
which are neighbours of a point which already is a member of S. To
make things simple we will assume that at the start, all the $q_{0}$
planes are saturated. (We do not want to adjust any of these planes)%
\footnote{To avoid `starting troubles', choose $q_{0}$ and $N_{0}$ s.t. $n<2^{q_{0}}$
and $N_{0}>n$%
}. Put Counter = 0.

Step 2: If no more points in G go to step 7, else: Randomly choose
a new point from G, which could be a candidate point to be put in
S, from the remaining points not in S, go to Step 3.

Step 3: Check if this new point is in a new `quadrant', this involves
finding its Orientation Vector and comparing with those in S. If the
point is in a new quadrant, put this point in S and store its Orientation
Vector in V, put $N=N+1$ go to Step 2, if not, it means it has a
neighbor in its quadrant. Go to step 4.

Step 4: (You will come here only if the current point has a neighbor
in S. Notation and procedure for this Step: We keep count of the number
of members of S which have first neighbors. Such points are called $a_{i}$,
its first neighbor will be called $b_{i}$, if $a_{i}$ has a second
neighbour this will be called $c_{i}$ and the third will be called
$d_{i}$ .) First: Find the `distance' of this new point from its neighbor, call this$\delta$ \textbf{If} $\delta < \delta_{th}$
then remove the new point from G itself and put it in set D, and Go To Step
2; \textbf{else} Put this new point in Set T. Check if this new point is in a quadrant which has already a pair in T, if not, put $Counter=Counter+1$ define $i=Counter$, call the point with which this new point is a neighbor as $a_{i}$. However, if this new point already is a neighbor of some point $a(j)$ already in S and if $b(j) \& c(j)$ exist call this new point $d(j)$, if only $b(j)$ exists call the new point $c(j)$ go to Step 2. However, if  $b_{j}$ , $c_{j}$ as well as $d_{j}$ exists this means the new point will become a 4th neighbor of $a(j)$,  a situation which we do not permit, so put back the new point in G and go to step 2. If $c_{i}$ and if $d_{i}$ do not exist, it means this point $a_{i}$ has only the present point as neighbor. 

  Calculate the mid point of the \textquotedbl{}segment\textquotedbl{}$(a_{i},b_{i})$,
if not already calculated, and call it $m_{i}$, add the coordinates
$m_{i}$, in a {}``List of Midpoints''.( Note $Counter$ only keeps track of the first neighbors of the the $a's$.)

If $Counter=n$ go to Step 5, if less than n, go to Step 2.

Step 5: Since $Counter=n$, this means you have collected n points
in T, each of which have to be separated from their neighbors. However
we, first re-check if all the $n$ points collected in T are in different
quadrants (This step is only cautionary, and not necessary if step 4 has been done properly),if say $b_{i}$ and $b_{k}$ in T
are in the same quadrant, then mark one of them say $b_{k}$
as $c_{i}$, if the latter doesn't exist, else as $d_{i}$ (If both
$c_{i}$and $d_{i}$both exist there is no place for $b_{k}$ put it
back in G , restore the count of points in G as well as restore in
either case$Counter=Counter-1$, check similarly for other $b's$
and then go to Step 2.)%
\footnote{We have permitted three points in T to belong to the same quadrant
but not 4 or more. This is just to simplify the algorithm, anyway,
such events are extremely rare. Since, for large $n$, if there are
$q$ planes there are $2^{q}$ quadrants, hence the chances of two
points randomly chosen to be in the same quadrant is $O(1/2^{q})$.
We permit max. no of neighbors,3, just to ensure that even in the
freakiest conditions the algorithm comes to a halt successfully. Of
course the the other possibility is the presence of points of `accumulation',
however this does not happen when points represent integers.%
}

Now in this step we will separate the n pairs collected in T by introducing
a new plane which passes through all the $n$ mid points, whose coordinates
are available in {}``List of Midpoints'' in the array M: {}``List
of Midpoints'' and determine the coefficients of the new plane, by
solving the$n$ constraint eqs. to pass it through the $n$ midpoints;%
\footnote{The coefficients of the plane containing the n midpoints, can be found
by using Gauss elimination, Gram-Schmidt evaluation or by using the
QR algorithm, the last is more useful just in case the n mid points
fall in a plane of n-1 dimension or less, then the rank of the coefficient
matrix will become less than n. This will happen rarely, and even
if it does, it only means we can accomodate another point (or points)
in T with a neighbor in S, thus making the rank n. In such a case
all the points in T (even though they are now more than n) can be
separated by a single plane - the algorithm can proceed.%
} call this plane as plane number $q+1$ and put $q=q+1$, and include
this new plane in S along with all the $n$ points labeled $b_{j},j=1,2..,n)$
and their coordinates, put $N=N+n$,

Step 6: Now check the neighbors of all the $a_{i}$ ,$i=1,2,..,Counter$, in T,
if only $c_{i}$ exists then it gets promoted to first neighbour.
But there is a slight complication after $a_{i}$ and $b_{i}are$
separated $c_{i}$ can be a neighbor either of $a_{i}$ or $b_{i}$
so if it is the former call $c_i$ as $b_i$ else call $b_i$  as
$a_i$ and $c_i $ as  $b_i$ the same kind of investigation needs
to be done for $d_i $ because it can now belong to the old  $ a_i $ 
quadrant or belong to the new $b_i $ quadrant, in either case the
$d_i$ gets promoted as second neighbour and it will be called $c_{i}$.
Calculate the new mid point $m_{i}$ for the just created segment
$(a_{i},b_{i})$ and store.

After finishing this task, the number, $r$ of second neighbours that
were promoted as first neighbours will be known, after drawing the
plane $q+1$, then call $Counter = r$.

Update V, the Orientation Vectors of points now stored in S wrt to
this new plane (their dimensions become $q+1$); Clear the data in
M: {}``List of Midpoints'' of all points T you have transfered to
S and remove data, of these $k$ points that have been transfered
to S, from set T (most of the time $ k=n$) and go to Step 2.

Step 7: (You will come here only if no more points are left in G $N_{f}=0$
and Counter is not yet $=n$ ) If Counter = 0, Stop else introduce
a new plane passing through the midpoints of segments collected so
far, (T will contain as many points as the value of Counter), since
Counter $<n$, some of the coefficients of this plane can be randomly
chosen; (take action like step 6 to check all these new points are
in their own quadrants), update V, the Orientation Vectors of points
now stored in S wrt to this new plane (their dimensions become $q+1$);
call this plane $q=q+1$, Counter = 0, Clear the data in M: {}``List
of Midpoints'' then if now $N_{f}=0$ i.e. G is empty, Stop; else
go back to Step 2. \textbf{At the end of section 6.2, we mention that
the case when $Counter<n$ and G is empty which is one of the possibilities
in Step 7; the situation can be dealt with in an easier manner.}

END OF ALGORITHM

The algorithm works swiftly most of the time, since in our case all
the point are discrete and there are no limit points or ther is no
{}``accumulation points'' in G the algorithm will always come to
a halt successfully. The general case when the input data is say prepared
by another program , then certain conditions need to be met to ensure
the correct halting of the algorithm. These are idscussed in the cted
reference.

\subsection*{DISCUSSION:}

The following crucial concepts will immediately clarify the steps
of the whole algorithm:: In n dimension space, if we are given n line
segements;

$(a_{1},b_{1}),(a_{2},b_{2}),(a_{3},b_{3}),...,(a_{n},b_{n})$,

Then a single plane passing through the mid points of each segment
will separate the n points $a_{1},a_{2},a_{3},..,a_{n}$ from the
n points $b_{1},b_{2},b{}_{3},..,b{}_{n}$ that is the $a's$ and
the $b's$ will lie on opposite sides of the plane. \textbf{But this
does not ensure that the $a's$ are separate from each other and the
$b's$ are separate from each other.} In order to ensure this we have
imposed the condition that each of the $a's$ belong to Set S and
are \textbf{already separate from each other }and have all have different
Orientation Vectors (in the space of q planes which are in S), \textbf{Similarly
by imposing the condition that each $b$ , (each of which belong to
T), has only one neighbor in S}, (for the moment ignore the second
and third neighbours$c's$ and $d's$) we are making sure that each
$b_{j}$ is separate from other $b's$ though, each $b_{j}$ is not
separate from its neighbor in S namely $a_{j}$. Thus each pair of
points $(a_{j},b_{j})$ have the same Orientation Vector. Now if the
$(q+1$)st plane is drawn then it will ensure that all the $2n$ points
i.e. the set of all the $a's$ and the set of all the $b's$ are not
only separate but are separated from each other. We can then include
plane $q+1$ and the $n$ points viz. the $b's$ in S. This intuitive
result is mathematically easily demonstrable:\textbf{ }

\textbf{Proof:} Since we have included the $(q+1$)st plane, all the
Orientation Vectors have gained one more dimension and have $q+1$
dimensions, the last, $(q+1$)st component of the Orientation Vector
of point $a_{j}$ will differ from the last component of the Orientation
Vector of its ex-neighbor $b_{j},$ because they are now on either
sides of plane $(q+1$). Thus proving that all the $2n$ points are
now separate\textbf{. QED}

The rare cases of three points belonging to the same quadrant (viz
$b_{i},c_{i},d_{i}$ ) is permitted but not four, we do this only
to avoid unecessary return of points to G, once randomly chosen from
G. However, one can simplify the algorithm if one just returns the
second point to G and then choose a new point at random.%
\footnote{This simplification of the algorithm will work most of the time work.
But there may be some freakish cases when (say) the last few points
left in G all belong to the same quadrant! This would mean introducing
a few unnecessary number of planes in the end.%
}

\subsection{PROOF OF ALGORITHM:}

At the start S only contains points which are separable from each
other, because that is the way they have been selected. Since $q=q_{0}$
have been selected initially each of the Orientation Vectors of the
points in S, are $q$ dimension vectors. They are all different.

Now till the time we come to step 5 we are just collecting points
which are separable and putting them S or in case thay have a neighbor
in S then we are putting the points in T. This can go on till there
are n points in T and we arrive at Step 5.

At this point there are n points in T and N points in S. Firstly,
there can never be two points in S which are neighbors to a single
point in T. This is because every point in S has a different Orientation
(Hamming) Vectors so they belong to different quadrants in Hamming
space and one point in T cannot belong to two quadrants.

There are now only two possibilities (i) Each point in T has one distinct
neighbor in S or (ii) there are two or three points in T which have
the \textit{same} neighbor in S. We will consider the first possibility
(i): Since by choice, every point in T must have one neighbor in S,
they all belong to different quadrants, since there are n points which
can be separated by the $q+1$ st plane after implementing Step 5.
Then in the new $q+1$ space all the points (the old points in S and
the n new points in T) are all separable form one another and hence
all the n points in T which awere first neighbours the $b_{i}$'s
can now be included in S, we must add the $q+1$ st component to each
Orientation Vector in S to make them into Orientation Vector wrt to
the $q+1$ st plane.

Now coming to the second possibility, if there are two or three points
in T which have the same neighbor in S then after introducing the
new $(q+1)$ plane only one of the points in T namely the one which
is the first neighbor of $a_{i}$ say $b_{i}$can be transfered to
S and the other points $c_{i}and$ $d_{i}$ need to be kept in T with
one of the points $a_{i}$ or $b_{i}$ as its neighbor and the newly
calculated midpoint.(Because the $q+1$ plane has separated the $a_{i}$and
$b_{i}$ they are on opposite sides of plane $q+1$, so we have to
now determine on which sides $c_{i}and$ $d_{i}$ are; the seemingly
complicated arguments are only to ensure that we make the right decision
about $c_{i}and$ $d_{i}$). However, the situation (ii) is rare in
high dimension n space, since we are choosing points in random, and
can best be avoided by not choosing two points in T which are neighbors
to the same point in S (this means ejecting the second point from
T and putting it back among $N_{f}$ in G, or by keeping it in T and
use it, later, only after you have introduced a new plane or even
after a few new planes.) In brief, the situation (ii) can be avoided
though it just causes difficulties in programming.

So we see after step 5 and Step 6 we will have an enhanced set S with
more points and one more plane and all of which are separated by these
$q+1$ planes. So we can call $(q+1)$ as $q$ and then re-do steps
2 to 6 till all points $N_{f}$ in G are exhausted. Then S will contain
all the points and all the planes which separate them.

Step 7 will happen in the end when all the points are exhausted but
there are less than\textit{ n }points in T and these must be separated.\textbf{
}The logic of Step 7 is similar to step 6\textbf{.}

\textbf{It may be mentioned here that Step 7 can be completely avoided.}
If we find that there are say less than n points in T. Let us say
r is a number such $Counter+r=n$. All we have to do is choose r random
points (these should not have been in G) each of which has as a neighbor
in its quadrant, one of the points which was drawn from G but now
in S. Then put each of these r points along with its neighbor as a
pair in T. Since you have already $Counter$ number of pairs in T,
with the addition of these r pairs, we have $n$ pairs. Now since
$Counter=n$, we can choose the last plane separating all the $n$
pairs.%
\footnote{These are just `End-Game', tricks that can make programming simpler%
}

So we see at the end of the algorithm S will contain all the points
along with the equations of the $q$ planes that separate them,\textbf{
}which are the needed outputs. \textbf{QED} %
\footnote{It may be noted that this algorithm is non iterative, and every point
is mostly {}``looked at'' only once.%
}

In the above algorithm it is assumed that every new plane does not
`split' some other point (pass through). However if this happens (it
is a very rare event), to some point, we will have to shift the `mid
points' of the segments slightly to one side (because to cut a segment
into two parts it is not necessary that the plane passes exactly though
its mid point) so that the plane passes to one side of the offending
point, %
\footnote{Instead of shifting a point already in S, we could shift all the mid points
by a small distance $\delta$ in the direction of the new normal
and recalculate the plane,thus moving it to one side
of the point upon which it was formerly incident. We do this so the algorithm won't fail. 

Another neat trick
which, pure mathematicians who are disciples of Cantor, may admire, is to replace the constant 1 by $\tau$
 (this 1 occurs in every  equation
 which defines the q  planes in Eqs., (1), (2),..., (q) below); and choose $\tau$ to be a transcendental number which is almost unity say choose $\tau = \pi/ \pi_{20}$, where $\pi_{20}$ is the value of $\pi$ correct  to 20 decimal places.
Then we can be assured that the plane can never pass through a point which is represented by integer coordinates,
 this is especially true since all the coefficients $\alpha_{i,j}$ are rational, being obtained by solving a finite set of equations whose coefficients are rational (because all the mid points are means of two rational numbers). This replacement of 1 by $\tau$ (which is very close to unity) should be done only after we have obtained all the $\alpha_{i,j}$,  we thus doubly ensure that all points which represent prime numbers will never lie exactly on any of the q planes. Strange as this may seem we have, succeeded in separating all prime numbers in n-dimensional space by using q planes, which act as boundaries which contain no rational but only algebraic and transcendental points!} and then proceed with the algorithm.

\medskip{}

\subsection{Conditions for the Algorithm to Stop}

Before completing the proof of the algorithm, it better to briefly discuss
the conditions when the algorithm will stop and when it will not and
to make sure that it will always stop.

In order to better understand what is happening: We will pretend that
the first part of the \textbf{If...then} statement in Step 4 is removed.
ie the `distance' is not calculated and all new points which arrive
at this step are never removed and put in D even if it is close to
its neighbor. (This analysis will then reveal the need of set D).

It is then clear from the algorithm that as we randomly pick points
from G each point picked finds a new quadrant or finds itself in T.
If it finds itself in a new quadrant there is no problem: It is put
in S and the algorithm goes to Step 2, and a new point is picked from
G. But if it lands itself in T, then it is paired with another point
which is already in S. Let us say that such a point is Q and it has
been paired with P which is in S and we pair(P,Q). Now the algorithm
will not attempt to transfer Q to S from T until it has collected
$n$ such pairs. Then the algorithm draws a plane through the mid
points of $n$ pairs. Every thing is fine if all the pairs are in
different quadrants the new plane separates all the $2n$ points and
the $n$ points which were collected in T (and now separated)are transfered
to S. The algorithm ensures that in the set of $n$ pairs if there
are at most 3 ($r=3$) points which are neighbors of a point in S (in the same quadrant), for example $b(i),c(i),d(i)$ can be neighbor of $a(i)$ which is in S, then things would be fine.

Things will not be fine if every new point that is subsequently picked
will all belong to the same quadrant as Q! This strange phenomena
will happen when: 
\begin{enumerate}
\item All the Q's which are chosen all belong to a sequence which has an
{}``accumulation'' point. That is the remaining points in G which
are left after all the others are separated are only those points
which tend to be close to some unknown accumulation point (or limit
point) close to P. This will happen since we are treating X-Space
as real space and all the coordinates $(x_{1},x_{2},x_{3},..,x_{n})$
are either rational numbers (or even real numbers which are uncountable as opposed to integers which
are countable). And then we are trying to separate an infinity of points  or uncountable points in a sequence by a countable (or finite) set of planes an impossibility! 
\item When all the points in G are not distinct and there are repetitions
in data. 
\end{enumerate}
They way to get rid of both the situations is to take a bit of care
in preparing the input data. In real life situations the input set G may be points generated by another computer program. The following are obvious suggestions:
\begin{enumerate}
\item We have characterized all the components of the $n$ space as real
numbers, that is all the $x's$ in the coordinate array of point P
: $(x_{1},x_{2},x_{3},..,x_{n})$, are considered real. It is better
that one of the components is an unique integer number. For example,
in the case of the medical data which we describe in section 3.2 (below)
it is better to label each point with an unique integer number ie
every patient P is given a unique integer label $L_{P}$, and we put
this value as (say) the $10^{th}$ component of the coordinate of
P. That is we put $x_{10}=L_{P}$. If this happens no two points will
ever have the same coordinate, (because they differ in their $10^{th}$
component) and the number of points will always be an integer number,
and be kept finite hence countable. They will always be separable because the minimum
Euclidean (or Manhattan)distance would be 1.
\item For every prospective candidate for T, we check its `distance' $\delta$
from it neighbor and if $\delta<\delta_{th}$ , we remove the point
from G and put it into the `Dust-Bin' set D. We use a small threshold
value for $\delta_{th}$, we need only calculate the `Manhattan distance'
for $\delta$. This process removes all `accumulation' points from
the data and thus explains the need of the first part of \textbf{If...then}
statement in Step 4.
\item Ensure that there are no repetitions in input data i.e. the points
in G are all distinct.
\end{enumerate}
Once it is ensured that the input data, set G, is prepared as above,
the algorithm will run smoothly.

\subsection{Calculation of coefficients of planes}

As we have seen, as the algorithm progresses and points are being transferred from G to S and new pairs are being created, one will have to draw a new plane as soon as $n$ pairs are collected. Let these $n$ segments be called:

 $(a_{1},b_{1}),(a_{2},b_{2}),...(a_{j},b_{j}),...,(a_{n},b_{n})$,
 
 and let the mid points of each segment such as $(a_{j},b_{j}) $ be denoted as $m_j$ thus we have the list of midpoints for all the $n$ segments:
 
  $( m_1, m_2,...,m_j,...,m_n)$, where the coordinate of the $j^{th}$ mid point is denoted as $x_1=m_{j1}, x_2=m_{j2},...,x_i=m_{ji},...,x_n=m_{jn}  $
  
  Now we require that the $q'=q+1$ plane:  
 
\begin{equation}
1+\alpha_{q'1}x_{1}+\alpha_{q'2}x_{2}+....+\alpha_{q'n}x_{n}= 0 
\end{equation}
should pass through all the above $n$ mid points, we thus have the $n$ constraint equations, from $(j=1,2,...n)$ :
\begin{equation}
1+\alpha_{q'1}m_{j1} +\alpha_{q'2}m_{j2}+...+\alpha_{q'i}m_{ji}.+....+\alpha_{q'n}m_{jn}= 0 
\end{equation}

The above Eqs.(4)represent $n$ linear equations which can be solved by suitable numerical techniques, eg. Gaussian elimination, to obtain the $n$ unknown coefficients $\alpha_{q'i}$, for $(i=1,2,...n)$. 

It may be noted that since all the  $m_j$ are midpoints of the segments $(a_{j},b_{j}) $ and the coordinates of $(a_{j}$ and $b_{j}) $ are rational numbers the coordinates of $m_j$  are also rational numbers, this makes all the coefficients 
$\alpha_{q'i}$ rational numbers. This observation has profound implications as we have described in the footnote 9 at end of page 12.

\medskip{}
 \textbf{ In APPENDIX A we have worked out a simple example on the
working of the algorithm in great detail. It may be consulted by anyone
who needs to quickly get an hands on experience of the algorithm and
program it.} \medskip{}

\subsection{Calculation of Computational Complexity}

The computational complexity of the algorithm is easily determined.

\textbf{Given:} $N_{f}$: Total number of points; $q_{f}$ : Total
number of planes used; $n$; Dimension of X-Space.

1. To determine the coefficients of each plane by Gaussian elimination
we require $\frac{2}{3}O(n^{3})$ multiplications, and $\frac{2}{3}O(n^{3})$
additions therefore for $q_{f}$ planes we have $\frac{2}{3}.q_{f}.O(n^{3})$
multiplications and $\frac{2}{3}.q_{f}.O(n^{3})$ additions. 2. To
compute the Orientation Vector of each point with respect to $q_{f}$
planes require per point per plane: $n$ linear evaluations, each
consisting of $n$ multiplications and $n$ additions. Therefore Total:
$N_{f}q_{f}.n$ multiplications and $N_{f}q_{f}.n$ additions. 3.
When every time a new point is put into S, it's Orientation vector
is compared with the Orientation Vector of the points in S. Therefore
there are $N_{f}^{2}$ Hamming Vector comparisons. But just to see
if two Hamming Vectors of dimension $q_{f}$ are not equal it does
not require us to check all the $q_{f}$ components, the moment the
rth component differs the Hamming vectors are declared unequal. So
we may assume that 99 percent of the time only about 8 components
are checked on the average. So the number of bit comparisons made
are: $\sim8.N_{f}^{2}$ bit comparisons.

4. We must calculate the Manhattan distance of every candidate member
which arrives at T from its neighbor (Step 4). For every plane there
will be $n$ such points arriving at T. Each Manhattan distance calculation
requires $n$ subtractions and $n$ additions, therefore for $q$
planes we have $q.2n$ additions. We ignore the additions required
to detect accumulation points which we had put in the Dust-Bin set
D, assuming they are small in number.

Now we need to guess as to how many planes are require to separate
$N_{f}$ points, we use the estimate of Boland and Urrutia (1995),
Ref {[}2{]}, also see {[}1{]},and assume that $q_{f}\approx log_{2}(N)$,
hence we come to the conclusion that the order of computations involved
by the algorithm are :

$N_{f}log_{2}(N_{f})n+\frac{2}{3}log_{2}(N_{f})O(n^{3})$ multiplications;

$N_{f}log_{2}(N_{f})n+\frac{2}{3}log_{2}(N_{f})O(n^{3})+log_{2}(N_{f}).2n$
additions and

$\sim8N_{f}^{2}$ bit comparisons.

\section{An application to Data Retrieval}

We will now briefly describe how this algorithm can be applied to
data retrieval. Suppose one has $N$ points in a $n$ dimension X-Space
and the algorithm was used to separate all N points and it was found
after running the algorithm that $q$ planes were finally necessary.
We will now show that this information can be used as a data retrieval
device. That is all the data can now be stored in such a manner that
retrieval can be done with great efficiency.

\subsection{Application to Image Data}

Now for purposes of this illustration we will assume that each point
N represents $n$ dimension image (say we had used a passport size
photograph which represents a face using precisely $n$ pixels) Now
we wish to store these images in such a manner that retrieval becomes
easy. The idea is simple: after we have solved the problem, we will
have the exact information of how each of the data points $N$ reside
in X-Space with respect to each other and the separation planes, because
we have the Orientation Vector for each point stored in S. We can
use this information to (i) store the images in such a manner that
(ii) it is possible to retrieve it easily. What we mean is that after
the storage is done and if we are given a fresh (approximate) image
of a person whose face is stored in the storage receptacle in the
repository, it possible to use this new (approximate) image to retrieve
the stored image.

We show that if the new photo which is a point P in X-space, is close
to the original photo, say a point L stored in the receptacle then
all we need to do is calculate $Ov(P)$ and compare with $Ov(L)$
, we will show presently by using the neural engine in the picture
below the task requires only $q.n$ multiplication $q.n$ additions
to retrieve L. Fig Ref:fig-fig-a 

\medskip{}
\begin{figure}[htp]
 \begin{center}
\includegraphics[scale=0.40]{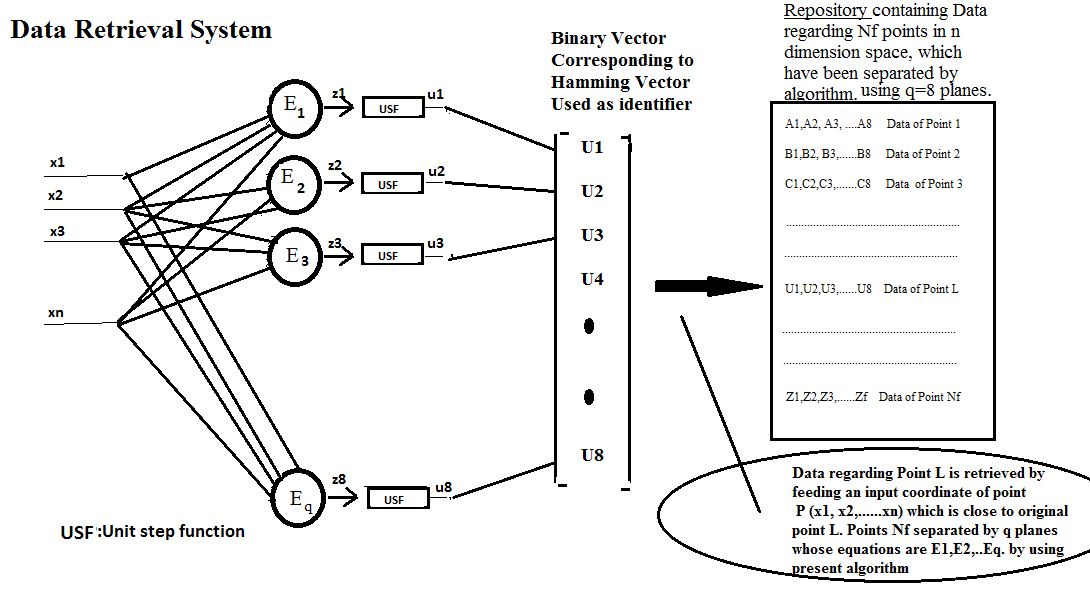}
\caption{An application of the algorithm for efficient storage and retrieval}

\label{fig:fig-a} 
\end{center}
\end{figure}

\medskip{}

In the figure $E_{1}$ represent the equation to the first plane stored
in S; We assume that the equation to this plane is gven by 
\begin{equation}
1+\alpha_{1}x_{1}+\alpha_{2}x_{2}+...,+\alpha_{n}x_{n}=0
\end{equation}
 we then define 
\begin{equation}
z_{1}=1+\alpha_{1}x_{1}+\alpha_{2}x_{2}+...,+\alpha_{n}x_{n}
\end{equation}
 If we use the Unit step function defined as: $Usf(z)$ defined s.t.
$Usf(z)=1$, if $z>0$ and $Usf(z)=0$ if $z\le0$ and define $U_{1}=Usf(z_{1})$,
then $U_{1}$ is 1 if the point P is on the +ve side of plane 1 and
is zero if it is on the -ve side of plane 1. the same goes for the
other planes $E_{2},E_{3},..E_{q}$. So the output array $(U_{1},U_{2},U_{3},...U_{q})$
is a binarized version of the orientation vector $Ov(P)$. Therefore
our {}``Storage Plan'' for each point L in is to use a binary representation
of the Orientation Vector $Ov(L)$ of a point $L$ as its label as
shown in the receptacle in the figure and then store information about
L in the space next to this code.

{}``Retrieval Plan'': Present an approximate image of L say P to
the network. It is then possible to immediately retieve the information
about L when a nearby point P has the same OV as L ie it is in the
same quadrant as L in X-space. Thus the coordinates $(x_{1},x_{2}...x_{n})$
of P when presented to the neural engine retrieves L. The retrieval
just takes $q.n$ multiplication $q.n$ additions.

\subsection{Application to Medical Data}

A similar application can be thought of in medicine. In this case
a person P may be represented as a point in n-dimensional feature
space and the $n$ numbers in the array $(x_{1},x_{2}...x_{n})$ may
represent, values of platelet count, WBC count, RBC count, MCV, MCH,
etc. a total of $n$ variables. So a doctor may be looking at patient
P, and wondering if there is any other person in the data base whose
medical condition is similar to that of the present patient P. All
the doctor has to do is to input $(x_{1},x_{2}...x_{n})$ to the Retrieval
system shown in Fig 1, and a L will be extracted s.t $Ov(L)=Ov(P)$,
hence it is very likely that P and L share similar ailments since
they share the same quadrant. The doctor can then examine the entire
case sheet of person L which is retrieved by the the repository shown
to the extreme right.

Thanks to the advancement of computer technology, algorithms such
as this one could be very attractive. In order to be aware that such
possibilities could soon become realities, let us obtain an approximate
estimate of the numbers involved: Taking the case of medical records;
assuming $n=50,N=10^{10}$ For such a problem since $n$ is large
it is reasonable to expect that approx. $q=40$, we see that the number
of calculations for finding these planes and separating all the N
points is approx $2\times10^{13}$ multiplications, Since computers
have achieved around 1 TFLOPS this translates to around 20 seconds.
To retrieve the data: for example the history of patient L, which
is similar to patient P, from a data base of 10 billion, would take
$q.n$ multiplications i.e. 2000 multiplications, which is practically
instantaneous.

\section{Properties and Interesting Aspects of the Algorithm}

In these two subsections we speak about one of the properties that
the algorithm has and some interesting aspects of the algorithm.

\subsection{The Algorithm follows Shannon's Principle}

Claude Shannon, held the view that a bit is an information and one
must use just the right amount of bits necessary to solve a problem.
(A mathematical Theory of Communication Bell Sys. Tech. J. pp. 379-423,
and 623-656, 1948).

We show below that the algorithm gathers and uses the information
optimally: We will define a completed `Stage' of the algorithmic process
as that state when a new plane say the $q^{th}$ has been inducted
into S along with the other points which it has just separated and
which are now in S along with all the Orientation Vectors of the points
now in S. To get to the next `Stage' the algorithm transfers enough
number of points and garners just sufficient information for it to
draw the $(q+1)^{st}$ plane. We will now prove that the new information
collected in bits is exactly equal to that which will be stored after
the next stage is completed that is after the $(q+1)^{st}$ plane
is drawn. No extra bits are collected, so the algorithm always makes
optimum use of information it collects `Stage' by `Stage' right up
to the end when the problem is completely solved.

Let us imagine: we are at a `Stage' when we had just added the $q^{th}$
plane which then separated all $N$ points in S, since we have calculated
all the $N$, Orientation Vectors w.rt. each of the planes we have
used $Nq$ bits of information to arrive at this `Stage'. Now let
us say as we proceed with the algorithmic process: We added $k$ new
points which happened to fall in empty quadrants and $n$ new points
which are paired with the old. When the $(q+1)^{st}$ plane is drawn
all these $(k+n)$ new points are added to $N$, to make the number
of points in S equal to $N'=N+k+n$. But since we need to determine
the Orientation Vectors of these new points wrt the $(q+1)$ planes
now in S, we would be acquiring $(k+n).(q+1)$ bits of information,
in addition,since we have to update the Orientation Vectors of the
old $N$ points wrt this new $(q+1)$plane, we have to add one bit
for each such old vector (because the old orientation vectors were
of dimension $q$ and now it has to be increased by one to accommodate
the new plane) so we have acquired $N$ bits. So the new bits of information
acquired is: $N+(k+n).(q+1)$. So if we add this to the old $Nq$
bits already acquired we have a total of $Nq+N+(k+n).(q+1)=(N+k+n).(q+1)$
bits which is exactly equal to $N'.(q+1)$ bits now stored as Orientation
Vectors by the $N'$ points now in S. So we see, that in the process
of acquisition and usage and then storage of bits of information there
has been no loss nor any redundancy. And even though the points in
G are picked at random, Shanon's principle has been followed, strictly,
in an optimal manner in going from $q$ planes to $q+1$ planes and
then onwards till the final $q_{f}$ plane is determined and drawn.%
\footnote{In the author's opinion, this one of the truly beautiful features
of this algorithm.%
}

The algorithm uses the geometrical properties of $n-$dimension space,
to judiciously choose a plane passing through $n$ mid-points simultaneously,
at an appropriate time, to make all this possible.

\subsection{Some Interesting Aspects of the Algorithm}

We will now speak about a few interesting aspects of the algorithm:

1. Once you have solved the problem, there is no need to start from
the very beginning, if at some later date you need to separate a new
set of $N_{k}$ points. All you have to do is include this new set
into set G (which is presently empty) and start running the algorithm
from Step 2. Then these new points will be inducted into S one by
one and new planes added when ever necessary, till all the points
are separated and G is once again empty.Also since for large dimension
$n$ we have the relationship $q=O(log_{2}(N)$, when you have solved
the problem of $N$ points with $q$ planes then if at some later
date the number of points become $2N$ then $q\rightarrow q+1$ only
and when $N\rightarrow N+4N$ then $q\rightarrow q+2$. This is a
huge advantage.

2. Suppose at some future time you need to increase the dimension
of data say from $n$ to $n+1$ ? Once again there is no need to worry,
assume that all the existing points have the same coordinates for
the first $n$ components but for the $(n+1)$ st coordinate we define
its value as zero. Do the same thing for the planes, the $(n+1)$
coefficient of each plane is defined to be zero. That is for any existing
planes we just define the coefficient $\alpha_{n+1}=0$ eg. see Eq(1).
Now all the data has become $(n+1)$ dimensional and there is no need
to change anything else and as and when new data points which are
now dimension $n+1$ comes in, we put them into G and the algorithm
can proceed from Step 2, just as before.

3. Both the points 1. and 2, above raises interesting possibilities.
Suppose each data point of $n$ dimension, represented an image of
dimension $n$, you can incorporate new images at any point of time.
In fact you can incorporate images of larger size.

4. Let us extend 3. a bit further: Now if you want to incorporate
a different kind of data which are $r$ dimensional, let us say they
are different because they are data pertaining to spoken words, again
there is no problem! Just increase the dimension of data to dimension
$n+r$, so each point is an entity in $n+r$ dimension space. The
old data will live in the first $n$ dimensions with their components
from $n+1$ to $n+r$ defined as zero; whereas the new data pertaining
to words will again be defined in this $n+r$ dimension space but
will have their first $n$ components as zero and have their components
from $n+1$ to $n+r$ as mostly non zero.

5. So we see new data can always be added not only of the same type
but of different types and the old data is never discarded only the
dimension of space increases and this can go on from generation to
generation. The set S will be the repository of all knowledge in perpetuity!
As they say Knowledge never dies!

\section{Conclusion}

In this paper we have reported the discovery of a new algorithm for
separating a given set of $N_{f}$ points by planes in n-dimensional
space. The algorithm is noniterative and will always halt successfully.
It has the property of restart, if new points are needed to be separated
the algorithm can continue from where it left off. And at some later
stage if the dimension of the data is increased ($n$ to $n+r$) is
increased the algorithm can still continue from where it left off,
(after some adjustments) and tackle the new data points which are
of a higher dimension. The algorithm is insensitive to the type of
data; the points may represent, images, words or any other type. It
can handle even the types are mixed as discussed in Section 4.

A rigorous proof has been provided in the body of the paper and a worked example is given in Appendix A.

 The computational Complexity is of $O(n.Nlog(N))+O(n^{3}log(N))$, where $N$ is the given number of points and $n$ is the dimension of space.

\textbf{An Informal Essay on the New Algorithm and its Future Applications is contained in Appendix B}

\section{Acknowledgements}

The author thanks the management of Sreenidhi Institute of Science
and Technology for their sustained support. He proclaims his grateful
thanks to his wife Suhasini, for her willingness to be a sounding
board on innumerable occasions and listen to his monologues without
which he does not think this work would have ever been done.

\section{References}

1. William B. Johnson and Joram Lindenstrauss: Extensions of Lipschitz
mappings on to a Hilbert Space, Contemporary Mathematics, 26, pp 189-206
(1984)

2. Ralph P. Boland and Jorge Urrutia: Separating Collection of points
in Euclidean Spaces, Information Processing Letters, vol 53, no.4,
pp, 177-183 (1995)

3. K.Eswaran:A system and method of classification etc. Patents filed
IPO No.(a) 1256/CHE July 2006 and (b) 2669/CHE June 2015

4. R.A. Fischer: {}``The statistical utilization of multiple
measurements'', Annals of Eugenics, 8, 376-386 (1938); also Annals
Eugenics, 7, 179-188 (1936)

5. P.C. Mahalanobis: {}``On the generalized distance in statistics'',
Proc. Nat. Inst. of Sc. India, 12, 49-55 (1936)

6. McCulloch, W. and Pitts, W. A: {}``Logical calculus of the
ideas immanent in nervous activity'', Bulletin of Mathematical Biophysics,
5:115\textendash{}133. (1943)

7. A. N. Kolmogorov: On the representation of continuous functions
of many variables by superpositions of continuous functions of one
variable and addition. Doklay Akademii Nauk USSR, 14(5):953 - 956,
(1957). Translated in: Amer. Math Soc. Transl. 28, 55-59 (1963).

8. Paul Werbos: {}``Beyond Regression: New Tools for Prediction
and Analysis in the Behavioral Sciences'', PhD thesis, Harvard University,
1974

9. Rumelhart, D. E., Hinton, G. E., and R. J. Williams: Learning
representations by back-propagating errors. Nature, 323, 533\textendash{}536
(1986).

10. J. Schmidhuber: Deep Learning in Neural Networks: An Overview.
75 pages, www.arxiv.org/abs/1404.7828 (2014).

11. Yoshua Bengio: Learning Deep Architectures for AI. Foundations
and Trends in Machine Learning: Vol. 2: No. 1, pp 1-127 (2009).

12. K. Eswaran: {}``On the storage and retrieval of primes using
n-dimensional geometry'', sent for publication.

13. J.C. Hawkins with S. Blakeslee: {}``On Intelligence'',
Publ. Henry Holt and Co. NY. (2004)

14. D. George and J.C. Hawkins: Trainable hierarchical memory
system and method, January 24 2012. URL https:/ www.google.com patents
US8103603. US Patent 8,103,603

\section{APPENDIX:A }

\section*{WORKED EXAMPLE}

\begin{figure}[htp]
 \begin{center}
 \includegraphics[scale=0.50]{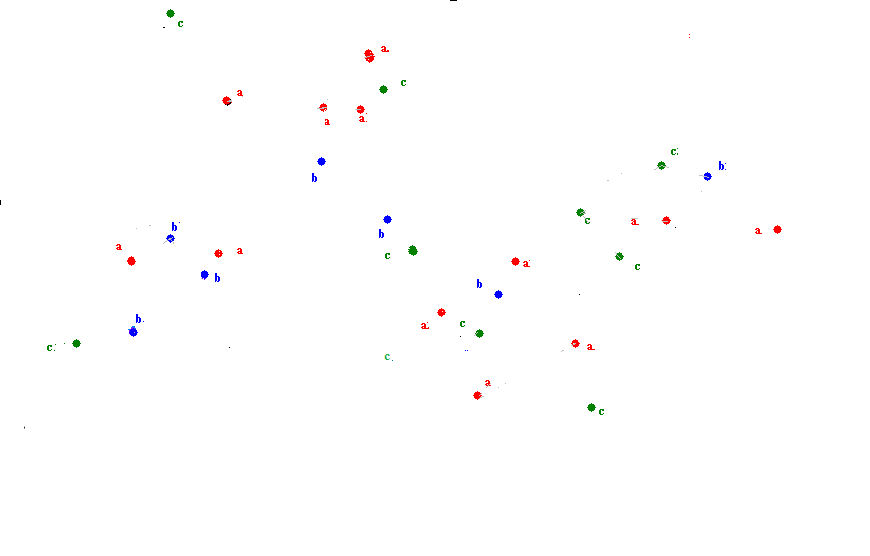}

\caption{Fig Shows the Example Problem which was used to demonstrate the working of our algorithm}

\label{fig:fig-b} 
\end{center}
\end{figure} 

\textbf{We will now show the working of our algorithm in complete
detail. The algorithm was used to separate the points shown in Fig } \ref{fig:fig-b}

It was discovered that the above points which are 29 in number in
2d space, can be separated by 8 planes, by using the algorithm. We
will now show how this was done. 

\begin{figure}[htp]
 \begin{center}
 
\includegraphics[scale=0.50]{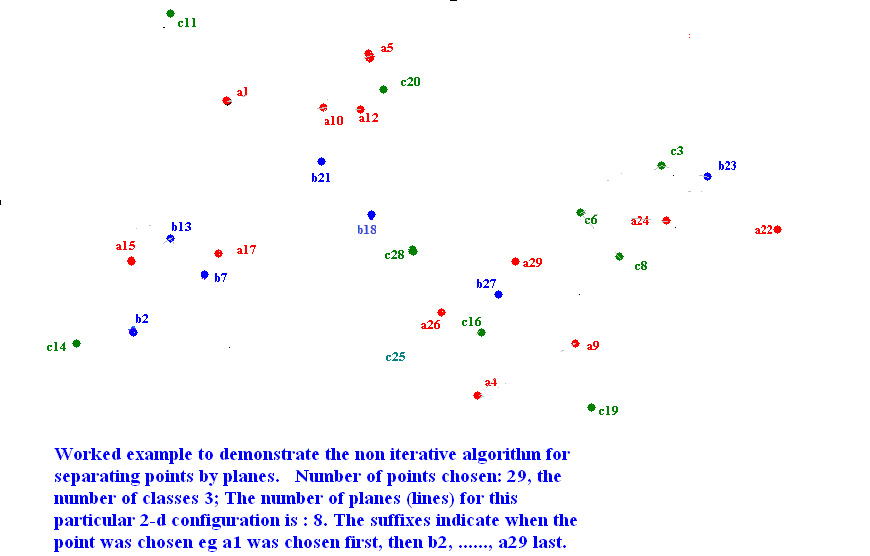}
\caption{Fig shows initial set of points to be separated.}

\label{fig:fig-c} 
\end{center}
\end{figure}

However, instead of using unlabeled points we will use Fig \ref{fig:fig-c} which
has a label for each point but in the same position. (Fig \ref{fig:fig-b} and Fig
\ref{fig:fig-c} are identical except for labels).

The 29 points belong to three classes, (class a,b,c). 

We have numbered each point by a number, these numbers is the sequence
with which each point was chosen to be in set  S (or T) when we solved
the problem. We now retain the labels while we describe the process
that was followed so that the method can be better understood. 

We describe the process in Stages.

\medskip{}

\medskip{}

\textbf{STAGE1:}

The first stage is when we started with an initial set of points which
are separated by planes.

In our commentary below we use the present tense in describing what
was actually already done, this change of tense from past to present
has been adopted so that the commentary reads better. 

\medskip{}
\begin{figure}[htp]
 \begin{center}
 
\includegraphics[scale=0.50]{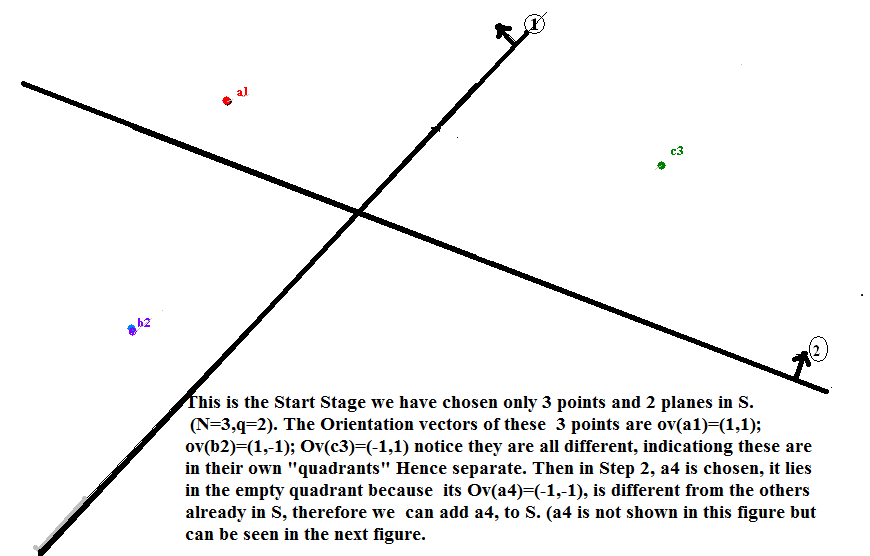}
\caption{Fig Start of the algorithm: Initial contents of S}

\label{fig:fig-i} 
\end{center}
\end{figure} 

This initial set of planes are two in number and numbered 1 and 2
as shown in Fig \ref{fig:fig-i}
below. And the initial set of points, three in number are
$a_{1},b_{2},c_{3}$. We store the coefficients of the two equations
for the planes 1 and 2 in S. The equations for the two planes 1 and
2 will be of the form :

\begin{equation}
1+\alpha_{1}x+\alpha_{2}y=0
\end{equation}
\begin{equation}
1+\beta_{1}x+\beta_{2}y=0
\end{equation}

We compute the Orientation Vectors of these three points which are
shown in Fig \ref{fig:fig-i}. and store in S and put $q=2$ and $N=3$.
The first component of the Orientation vector of $a_1$ is $+1$ because it lies on the positive side of the normal of plane 1, indicated by black arrow, similarly the second component is also $+1$ because $a_1$ is also on the positive side of plane 2, therefore we write $Ov(a_1)=(+1,+1) $, the other Orientation vectors of $b_2$ and $c_3$ are depicted in Fig \ref{fig:fig-i}. The first component of $Ov(c_3)$ is $-1$, because it lies on the negative side of plane 2 but its second component is $+1$ because it lies on the positive side of plane 2, hence $Ov(c_3)=(-1,+1)$, the OV of point $b_2$ is also shown in the Fig \ref{fig:fig-i}.
Notice all the three point have different Orientation Vectors and therefore they all are in separate quadrants, of course this is as it should be, otherwise they would not have qualified to be members of S. 
 
Now applying
Step 2 of our algorithm we choose point $a_{4}$, and then go to Step
3, and check whether this point is in the same quadrant as the other
three points, in order to do this we have to find the Orientation
Vector of this point: This is done by substituting the $(x,y)$ of
the point $a_{4}$ in Eq. (2), for plane 1, the lhs will evaluate
to some -ve value which will signify that $a_{4}$ is on the -ve side
of plane 1 , then we substitute $(x,y)$ of the point $a_{4}$ in
Eq. (3) which is the equation for plane 2, the lhs will again evaluate
to some -ve value which will signify that $a_{4}$ is on the -ve side
of plane 2. In this manner we obtain the Orientation Vector  of point $a_{4}$
as $Ov(a_4) =(-1,-1)$. 

In n dimension space this is the only way we can determine
Orientation Vectors. And determining an orientation vector of a point
in n-space where q planes are present would involve evaluating q linear
equations of type Eq.(1), and therefore would need a total of $q.n$
multiplications and $q.n$ additions.

 However when n = 2 (or 3) and
we have a diagram such as the one below present, we can write down
the Orientation Vectors by sight, we can see
that $a_{4}$ is on the -ve side of the plane 2 because it is on the
opposite side of the normal direction of plane 1, which is indicated
by the black arrow on the top of the figure, similarly $a_{4}$ is
on the -ve side of the plane 2 because it is on the opposite side
of the normal direction of plane 2, which is indicated by the black
arrow in the middle right corner of the figure, hence we can conclude, correctly, that $Ov(a_4) =(-1,-1)$. \textbf{From now on
we will write down the Orientation Vectors by sight. }

We now complete the process given in Step 3 with regard to this new
point $a_{4}$, we see that by comparing the Orientation Vectors of
this point $a_{4}$ with the Orientation Vectors of the points which
are currently in S, namely $a_{1},b_{2},c_{3}$ , and finding that
it is different we are sure that $a_{4}$ is in its own quadrant space
and thus point $a_{4}$ is added to S, and its Orientation Vector is
stored in V and we put $N=N+1=4$ and go to Step 2. We now see that
there are no more free regions where a point randomly chosen from G that is points from Fig \ref{fig:fig-c} can be put in S.
And hence when we go to Step 2 and choose a new point, we will be directed to go to Step 3, as we shall soon see in Stage 2.

\medskip{}
\medskip{}

\textbf{STAGE 2}

 We are now in Step 2, we randomly select a point say $a_{5}$ notice
it has a neighbor $a_{1}$ . This is discovered by sight, but in actuality 
if we are in $n$ dimension space, we calculate the Orientation Vector
of this new point, $Ov(a_{5})=(1,1)$ and compare with all other Orientation
Vectors of points already in S, namely $a_{1},b_{2},c_{3},a_{4}$
and then discover that $Ov(a_{1})=Ov(a_{5})=(1,1)$, this implies
that in Q-space $a_{1}$and $a_{5}$ are mapped to the same point,
hence in X-space they are not separated by any planes thus we discover
that point $a_{1}$ is a neighbor of the randomly chosen point $a_{5}$.

The above is the Standard procedure of discovering whether a new randomly
chosen point has a neighbor in S, this procedure is done in Step
3 for all randomly chosen points. \textit{This Standard procedure
of discovering neighbors in S, is assumed to be always adopted,} \textbf{
though for the purpose of this example we will from hence forth, do
the {}``neigbour detection'' in S, by sight. }Now that $a_{5}$ has
a neighbor we go to Step 4, calculate the mid point coordinate $m_{1}$ and
store it in the List of Midpoints and put $a_{5}$ in T. Put $Counter=Counter+1=1$
and then go to Step 2. We now choose a new point at random say, $c_{6}$
it has a neighbor point $c_{3}$ which is already in S, we find the
midpoint $m_{2}$ put $c_{6}$ in T and increase $Counter=Counter+1=2$, 
but this time we cannot go back to Step 2. Since we are in two dimension
space (n=2), we need to immediately separate the points in T from
their respective neighbors in S. As can be seen that the algorithm initiates procedures to start including a new plane, in S, as soon as $Counter=n$ (where $n$ is the dimension of Space).

\footnote{In n dimension space we could have gone back to Step 2 and collected
some more points in T, till Counter becomes n, indicating that n points
have been collected in T, each having one neighbor in S, and then
it will be time to separate the n points by introducing a new plane
which passes through the n mid points $m_{1},m_{2},...,m_{n}$ which
are stored in the {}``List of Midpoints''. The calculation of the
coefficients of this new plane is done in Step 5 and  the order of computations
to find the newplane (say) by using Gaussian elimination is $O(n^{3})$,
so if we have to intoduce $q_{f}$ planes in the problem the multiplications
are in the order of $O(n^{3}).q_{f}$.%
}
To proceed with our algorithm we need to go to Step 5. Step 5 says
that we must now find the equation to a new plane which passes through
the mid points $m_{1}$ and $m_{2}$, This is easliy done by assuming
that the equation to this new plane is :$1+\beta_{1}x+\beta_{2}y=0$;
with unknown coefficients $(\beta_{1},\beta_{2})$ then these two
unknown coefficients can be found by using the condition that the
line must pass through $m_{1}$and $m_{2}$, whose coordinates are
$(m_{1x},m_{1y})$ and $(m_{2x},m_{2y})$ to obtain the eqs:

\begin{equation}
1+\beta_{1}m_{1x}+\beta_{2}m_{1y}=0
\end{equation}

\begin{equation}
1+\beta_{1}m_{2x}+\beta_{2}m_{2y}=0
\end{equation}

Solving the above we determine $(\beta_{1},\beta_{2})$ which completely
defines the equation for plane 3,  which is now include S and  we drawn this line as shown in Fig \ref{fig:fig-d}. Since the coefficients $(\beta_{1},\beta_{2})$ are known the direction of the normal is also known, the normal is also drawn (black arrow) in the figure.

Now $q=q+1=3$. Going to Step 6, we check to see if all the points
are in their own quadrants (ie are separated) by actually updating
the finding the Orientation Vectors of the old points wrt to the 3
planes. The Orientation Vectors are now of dimension 3 and given by:

$ov(a_{1})=(1,1,-1); ov(b_{2})=(1,-1,-1);$ $ov(c_{3})=(-1,1,1);$
$ ov(a_{5})=(1,1,1); ov(c_{6})=(-1,1,-1)$.

We see that all the Orientation Vectors 

of $a_{1},b_{2},c_{3},a_{4},a_{5},c_{6}$ are all different so that
they are all in their own quadrants in Q-space and hence separated
from one another in X-space. This ofcourse, is plainly visible by
sight in Fig \ref{fig:fig-d}. 

Now we complete the tasks indicated in Step 6. we clear the ``List
of Mid Points'', include the new points $a_{5},c_{6}$ in S, include
plane 3 in S put q=3, Clear the set T because its contents $a_{5},c_{6}$
have been transferd to S. Then we go to Step 2. The situation arising
is shown in Fig \ref{fig:fig-d} below: 

\begin{figure}[htp]
 \begin{center}
 
\includegraphics[scale=0.50]{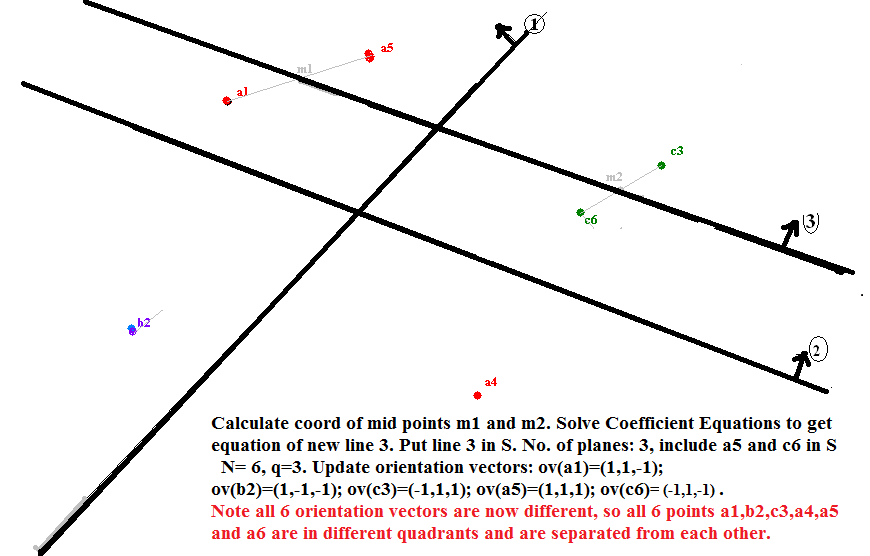}
\caption{This is the situation after completing  Stage 2}

\label{fig:fig-d} 
\end{center}
\end{figure}

\medskip{}

\medskip{}

\textbf{STAGE 3: }

\begin{figure}[htp]
 \begin{center}
 
\includegraphics[scale=0.50]{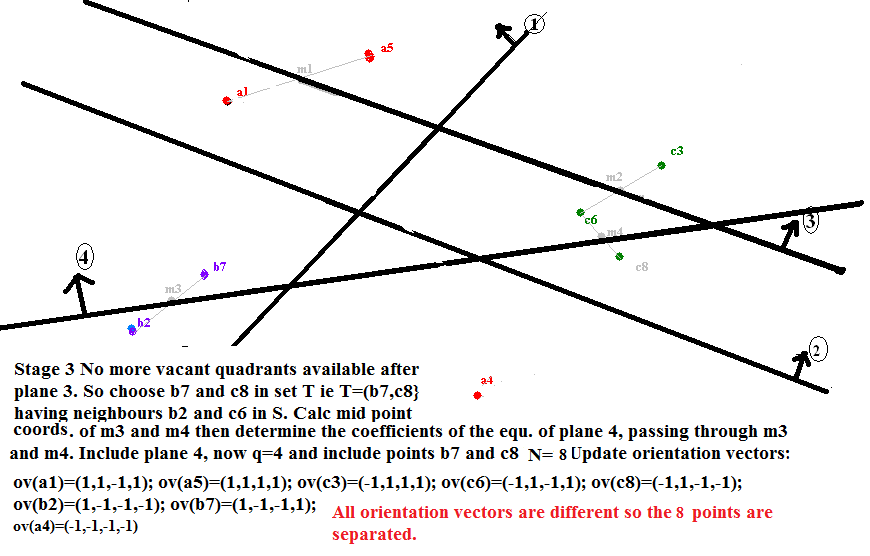}
\caption{This is the situation after completing  Stage 3}

\label{fig:fig-e} 
\end{center}
\end{figure}

We are now in Step 2. Since no more points can be added,the situation
is exactly the same as that in the beginning of Stage 2, so we repeat
what was done in stage 2 except that we randomly choose two new points
$b_{7}$ and $c_{8}$ which have neighbors $b_{2}$ and $c_{6}$ which
are already in S. So we follow the steps similar to that described
in STAGE 2 and end up with a new plane, viz. plane number 4, put q=q+1=4; and S
now contains 8 points all their Orientation Vectors in the new 4 dimensional
q space are all different and therefore all the 8 points are separated
by these 4 planes. See fig \ref{fig:fig-e}.

\medskip{}
\medskip{}

\textbf{STAGE 4;}

From now on we will hurry through the steps.

We are in Step 2. Two new points: $a_{9}$ which is neighbor of $a_{4}$ and
$a_{10}$ which is a neighbor of $a_{1}$ are introduced and a new
plane plane No. 5 is introduced and transfered to S.

\begin{figure}[htp]
 \begin{center}
 
\includegraphics[scale=0.50]{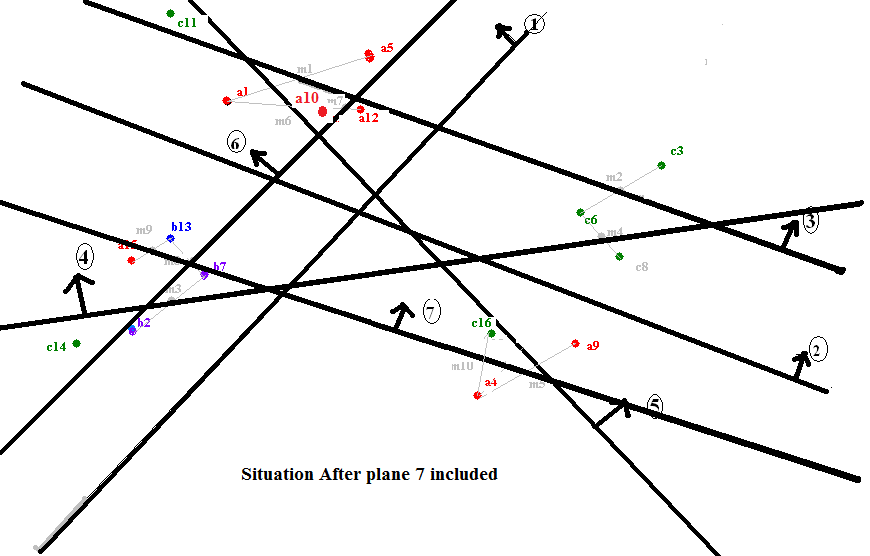}
\caption{Situation after planes 5,6,and 7 are introduced in Stage 4}

\label{fig:fig-f} 
\end{center}
\end{figure}

After this we see that the next point $c11$ occurs in an empty quadrant
hence it is immediately added to S, The next two points $a_{12},b_{13}$
have neighbors $a_{10}$ and $b_7$ resp. and therefore need a new plane, viz plane 6, and are all transfered to
S along with plane 6. The next point $c_{14}$ happens to be in its
own quadrant so is included in S. The next two points $a_{15},c_{16}$
have neighbors $b_{13}$ and $a_4$ in S and determine a new plane 7 and are transferred
to S along with 7. See Fig \ref{fig:fig-f}

\medskip{}
\medskip{}

\textbf{STAGE 5:}

We are in Step 2: 7 planes have been just added and S contains 16
points. All the Orientation Vectors are now dimension 7. They are
all different for the 16 points, as may be checked tediously or by
sight.

\begin{figure}[htp]
 \begin{center}
\includegraphics[scale=0.50]{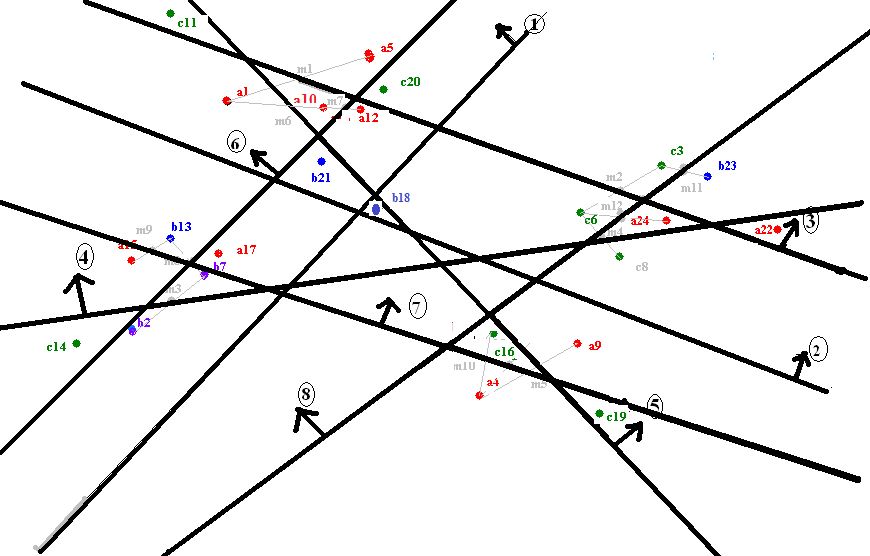}

\caption{Eight points and one plane are introduced in Stage 5}

\label{fig:fig-g} 
\end{center}
\end{figure}

The new randomly chosen points $a_{17,}b_{18},c_{19},c_{20},b_{21},a_{22}$
all are in their own quadrants and can be directly included in S with
out troubling ourselves to add a new plane see Fig \ref{fig:fig-g}. (This sort of situation
when we can include a whole sequence of points before we introduce
a new plane occurs when the number of planes q increase, because
the number of quadrants in q space increase. For q planes we have
$2^{q}$ quadrants in Q space, therefore there is space for more points
in q space.)\footnote{ The situation in large $n$ dimension space is much easier, this is because whenever you introduce a plane say from $q$ to $q+1$ the number of quadrants in X-space double from $2^q$ to $2^{q+1}$, this doubling in X-space stops after $q=n$ after which you will get some confined regions. Now an interesting question arises: If have just added a plane (say) the $q^{th}$ and you already have N points in S, how many points can you add to S before you need the next plane? The answer for large $n$ is that you can on the average add $O(N)$ points.}
 After this we arrive at $b_{23}$ and $a_{24}$ which
have neighbors already in S namely $c_{3}$ and $c_{6}$ . These determine
the last plane plane 8. See Fig \ref{fig:fig-g}

\medskip{}

\medskip{}

\textbf{STAGE 6:}

We are in Step 2 and have just introduced plane 8. There are 24 points
in S and 8 planes.

\begin{figure}[htp]
 \begin{center}
\includegraphics[scale=0.50]{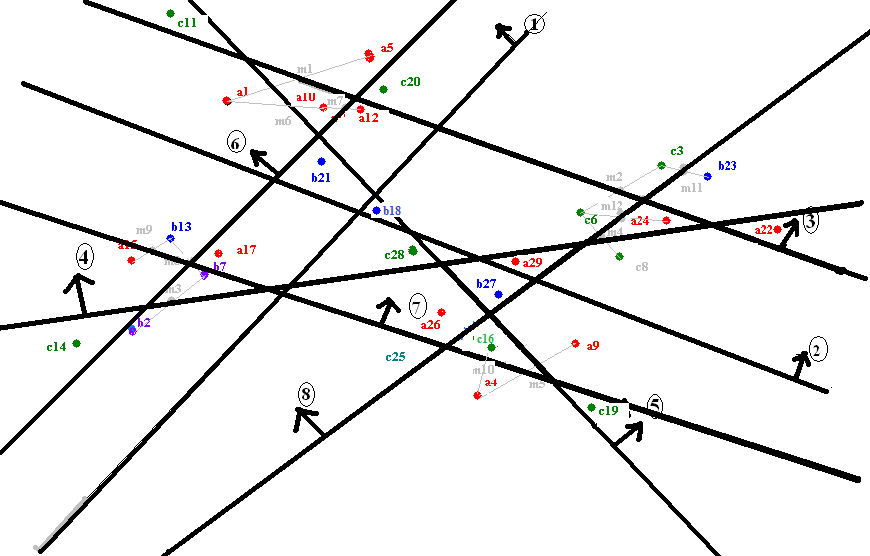}

\caption{After Stage 6 the final configuration all 29 points separated by 8 planes.}

\label{fig:fig-h} 
\end{center}
\end{figure}

We see that we now randomly choose the sequence of points $c_{25},a_{26},b_{27},c_{28}$
and $a_{29}$ all of them are in their own quadrants and are already
separated from other points in S and from each other and hence can be
included in S without adding any more planes. And now there are no
more points in G and so we have $N=29$and $q=8$. The Final solution
is given in Fig \ref{fig:fig-h}

\medskip{}

\newpage
\subsection{Tackling New data}

In this subsection we will, for the sake of completeness, tackle some of the situations that we had
discussed before, in section 4,but with reference
 to this \textbf{Worked  Example.}

How shall we tackle new points?  Suppose we have once solved a problem that is, we have separated $N$ points given in an original set G, in n dimension space and have found that $q$ planes have done the job. After this we make a retrieval and storage engine as depicted in Fig \ref{fig:fig-a}  and discussed in Section 3. We begin to use the retrieval engine for some time, and afterwards we encounter new data. 

\textbf{Case 1: New Data has the same dimension as the old data} 

\begin{figure}[htp]
 \begin{center}
\includegraphics[scale=0.50]{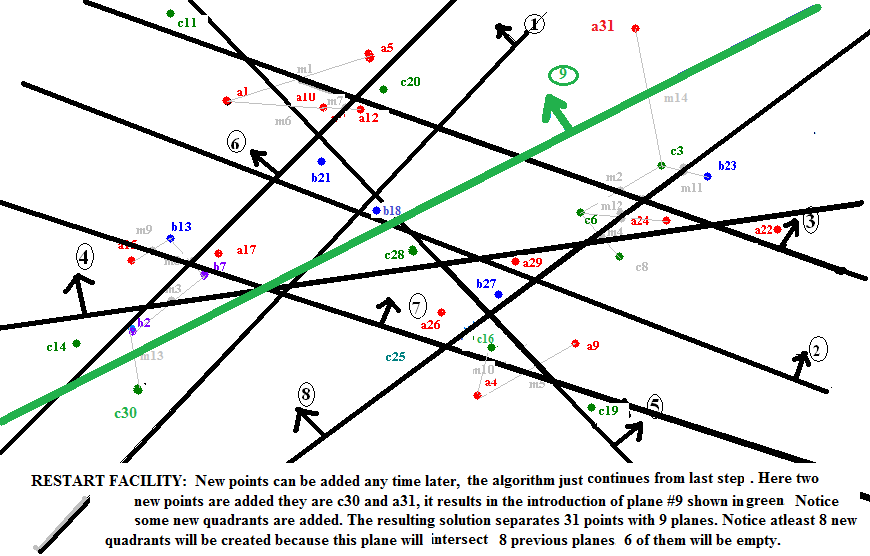}

\caption{Two New Points encounered: Resuming from last step.}

\label{fig:fig-j} 
\end{center}
\end{figure}

We have already spoken about this situation: We had said that there is no need to start from the very beginning, The algorithm can start from where it left off.We just add these new points to Set G, which until now was empty, and start the algorithm from Step 2 keeping the Set S containing the latest number of points and planes which we had (i.e. $N=N_f$ and $q=q_f$, with all the $Ov's$), just before we encounter this new data. We already depicted the situation, in our example, Figure \ref{fig:fig-h} , we have separated 29 points with 8 planes. Now we encounter 2 more points (say) $c_{30}$ and $a_{31}$. We just continue as before (assuming just for convenience that both these have neighbors): treat these two as temporary candidates in T with neighbors as shown in \ref{fig:fig-j} and draw a new plane number 9. This separates all the 31 points and  we have an additional plane, and these two new points all of which now can be included in S. Notice new quadrants are created, these new quadrants could capture some new points,  which are are still in G. The algorithm continues from Step 2 till G is empty and then Stops. If G becomes empty at $N=31$ then the situation in S is as shown in Fig \ref{fig:fig-j}.

\medskip{}
\medskip{}

\medskip{}

\textbf{Case 2: New Data does not have the same dimension as the old data, but has one more dimension $n=n+1$} 

Suppose we are as before in a situation depicted by Fig. \ref{fig:fig-h} i.e. S contains 29 points and 8 planes. Now we encounter three new points of dimension $n=n+1$, since all  our original data is 2d we can assume that all the points are in the X-Y plane. We assume that these three new points belong to the new data set and are of dimension $n+1$. 

We tackle the problem by first converting all the 2-d data in S to 3 d data, this is simply done by embedding the 2d data in 3d space. With reference to Fig \ref{fig:fig-c}

Step C1: Convert all the 2-d coordinates by defining for each point whose original coordinates are $(x_j,y_j), (j=1,2..N)$   a new coordinate   $(x_j,y_j,z_j), (j=1,2,...N)$  such that $ (x_j=x_j; y_j=y_j,  z_j=0) , (j=1,2...N).$

Step C2: Convert all the 2d equations of the $q$ planes to corresponding equations valid as 3d equations eg if the equation of the $2$nd plane is (say):

\begin{equation}
1+\beta_{1}x+\beta_{2}y=0
\end{equation}
we should convert it to : 
\begin{equation}
1+\beta_{1}x+\beta_{2}y + \beta_{3} z = 0
\end{equation}
and define 
\begin{equation}
 \beta_3 = 0
\end{equation}
 this process should be done for all the $q$ planes.

By this process a 2d plane will become a 3d plane, however its normal will be in the X-Y plane. If we assume that X-Y plane to be `horizontal' then all the planes will become vertical walls containing the original 2d lines. These planes are now shown as blue lines. All the original quadrants at this stage will become 3-d regions defined by vertical walls shown as blue lines. 

\medskip{}
\begin{figure}[htp]
 \begin{center}
\includegraphics[scale=0.50]{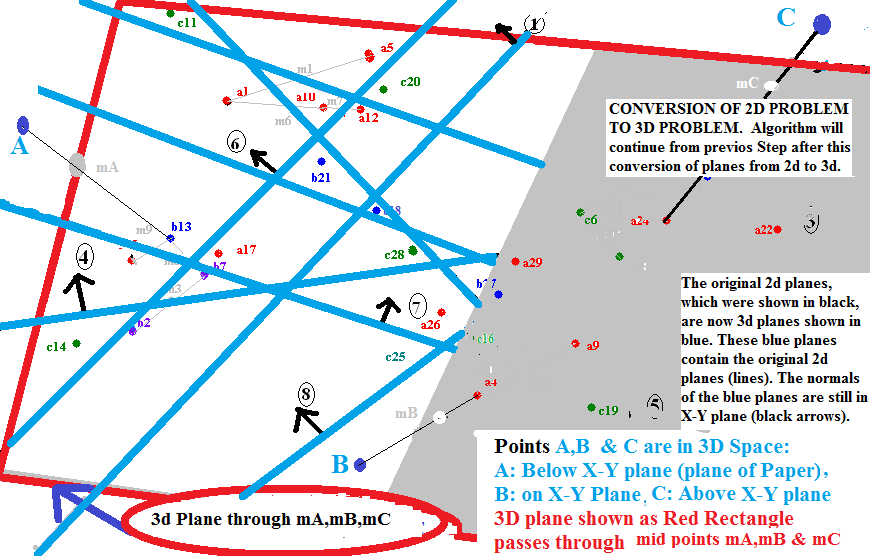}

\caption{The dimension of Problem has increased by one to n=3}

\label{fig:fig-k} 
\end{center}
\end{figure} 

We start the algorithmic process: We go to Step 2 of the algorithm after putting $Counter=0$ and $n=3$ and putting A,B,C in G. And S will contain $q$ 3d planes and N points along with their `Orientation Vectors'. For simplicity we will assume that none of the three points A,B,C fall in a new `quadrant'\footnote{If they do just transfer them to S} so they will all end up in T and each with their respective neighbors for example the neighbor of A and B are shown to be : $b_{13}$ and $a{4}$ resp. We will have three mid points which we call $m_A, m_B, m_C$. Now since $Counter = 3 = n$, we go to Step 5 and find the equation to the plane which passes through these three midpoints. The new `genuine' 3d plane\footnote{What we mean by `genuine' is that this $(q+1)$st plane is the first plane among all the planes in S, whose normal is not lying in the X-Y plane.  Of course as we proceed with the algorithm and be adding points in 3d space there will be many such planes in S.}   along with the old 8 planes will now separate all the N=29 points from A,B,C and from each other. \footnote{Here we assume for simplicity that no two points of A,B and C have the same neighbor.} So we can now add A,B,C to S and include this new plane so  $N=32$ and $q=9$ and calculate all the Orientation Vectors of the 32 points (i.e. update V) and then go to Step 2 of the Algorithm after clearing T and the `List of Midpoints'.

 So we see now S has 32 points all separated by 9, 3d planes if there are more points in G, the algorithm proceeds or else stops. Fig \ref{fig:fig-k}

We have thus seen how the algorithm can restart from the last step made earlier, even when the dimension of the new points are increased by one.

\textbf{END OF WORKED EXAMPLE}

\section{APPENDIX  B : An Informal Essay on the New Algorithm and its Future Applications}

\textbf{\large What the Algorithm Does}: Imagine we are given a set
$G$ of N points in n-dimensional space. Basically the algorithm finds
planes, in n-dimension space such that they can separate all the N
points, in such a manner that every point is separated from every
other point by at least one plane. The output of the algorithm is
a Set S containing the points and also the equations of the planes
that separate them. (In general for large dimension space the number
of planes $q,$required is approx.$log2(N)$).

\medskip{}

\textbf{\large How it\textquoteright{}s Done:} Let us imagine the
Set $G$ as an n-dimensional $X-$Space, containing $N$ points. We
create another n-dimensional $X$-Space called $S$. We then transfer
points randomly from $G$ to $S$ one by one, so that they occupy
the same coordinate position in $S$ as they had occupied in $G$
(their coordinates do not change) and also planes are drawn in S.
The algorithm makes sure that after a new plane is drawn, all the
points in S at this stage are separated. The algorithm proceeds Stage
by Stage transferring, new points from $G$ and drawing new planes
in S till eventually S contains all the N points as well as the q
planes needed to separate all the N from one another. 

\medskip{}

\textbf{\large Beauty: }There is a certain beauty in the algorithm,
there are no redundancies and duplications. (Ex. if it required say
97 bits of information to draw a plane, then it will acquire exactly
all these 97 bits of information and draw the plane, then store the
97 bits, before it seeks more information to draw another plane),
Sec 4.1, p. 15, contains the proof. It adopts Shannon\textquoteright{}s
principle that each bit is an information, so use it as far as possible,
before you seek another bit of information, (but these are technicalities
that I will pass on for some other time).

\medskip{}

Just to put the paper in its proper perspective, we list its contents: 
\begin{enumerate}
\item It has a very strict mathematical proof, of how $N$ points in n-dimensional
space can be separated by q planes.
\item The Complexity is of $Nlog2N$. This itself is somewhat unique because
very few algorithms have this efficiency to name a few:(i) the FFT,
complexity: $N.log(N)$, (ii) Euclidean algorithm of GCD complexity:
$log(N)$, $N$ being the larger of the two numbers, (iii) Quick Sort
algorithm complexity on the average $N.log(N)$ worst case $O(N^{2})$,
(iv) Gauss elimination for solving $N$ linear equations $O(N^{3})$,
(vi) Primality testing algorithm of Agarwal et al, Complexity $O((logN)^{6})$,
(vi) RSA algorithm $O(N^{3})$, here $N$ is the number of bits.
\item It contains the conditions under which the algorithm can be made to
work along with a proof. 
\end{enumerate}
We have Simple Worked example in the Appendix A and outlined applications
in Sec 3 and its interesting properties in Sec 4.

\medskip{}

\section*{A brief essay on the origins and relevance of the work}

\medskip{}

\subsection*{About separation of clusters vs. the separation of points }

Many researchers, from statisticians and neural network scientists
have long been trying to separate clusters by planes or discriminant
functions. This has been ever since the time of Fisher {[}4{]} and
Mahalanobis {[}5{]} and McCulloch and Pitts{[}6{]}, Kolmogorov {[}7{]},
Werbos{[}8{]}, and Rumelhart, Hinton and Williams {[}9{]} and others
{[}10{]}, also the Deep Learning people{[}11{]}, famous names in the
fields. But all have had great difficulties in large n-dimensional
space and they found that it is not easy. So the phrase like \textquotedbl{}The
Curse of Dimensionality\textquotedbl{} and \textquotedbl{}NP Hard
Complexity\textquotedbl{}, has become a part of folk lore. However,
from the very beginning it was never very easy to separate clusters
by planes. This is mostly because, a cluster is NOT well defined,
every cluster has its own shape and in $n$-dimensions you could have
long thin filaments and all kinds of snake like dragon like shapes
which constitute a cluster. Though statisticians try to approximate
the shape of each cluster as ellipsoids or even simple spheres, clusters
in general would require more parameters to define their shapes than
that required to define the planes which are supposed to separate
them! So all along it was, perhaps, very naive of all of us to have
tried to separate clusters when such entities are not mathematically
well defined. I felt that it is far better to separate the individual
points and to use the enormous space and degrees of freedom that is
available in $n$-dimension space to separate each point, rather than
try to separate clusters which will never be well defined. This was
the genesis of the idea that gave an impetus to do the kind of research
work reported in this paper. As an illustration, in the last section
we have considered a cluster of 29 points in 2-d, see Fig 3, 
and used the algorithm to separate all the 29 points using 8 planes
see Fig 9, (it can be proved that the theoretical minimum for
29 points in 2-d, is 7 planes).

Another fact, that only adds to the prospect of success in this new
direction is that: for large n dimension space where $2^{n}>N$ ($N$
being the number of points), you will find the number of planes $q$,
needed for separating $N$ points is far less than the number of points
itself, in fact $q=O(log2(N)).$

Imagine: Even a highly reduced small passport size image of $30X30$
pixels is a point in a $900$ dimension space. And in such a space
there are $2^{900}$ quadrants i.e. approximately $10^{270}$ quadrants.
And even if you put one image in one quadrant (thus automatically
separating them from others) you will never be able to fill up this
space: There are only about $10^{85}$ atoms in the universe so how
will you make so many photographs?

So you see this paper makes the {}``Curse of dimensionality'' into
a very Great Boon and Blessing! It is this aspect which has induced us to
 write this Short Note.

You may ask: Why did not the author speak about all this, in the main
body of the paper?

The answer is: We wanted all the readers to concentrate on the algorithm
and its proof and not rile them with matters not germane to the task
on hand; for after all a mathematical paper makes its greatest impact
by cold logic and rigid proofs rather than tall talk and philosophy.
\ldots{}.. 

In Sec 3.2 where we describe a possible application to medical records
of 10 billion people- the number of planes, $q$, you require would
be $30$ or $40$ planes. Most of the time you will require very few
planes to separate all$N$ points. This is counter intuitive but will
always happen, for large $n$ the number of planes $q$ will be $O(log2N)$.
The following `explains the phenomena\textquoteright{} : Consider
an empty $n$-dimensional space, void of planes, then if you put one
plane, it divides the space to 2 `quadrants\textquoteright{}, the
2nd plane will divide the space to 4 `quadrants\textquoteright{},
the 3rd to 8 and the qth plane to$2^{q}$ `quadrants\textquoteright{}.
(The doubling stops only when $q=n$ , afterwards some closed regions
are formed), so you see you have sufficient number of planes $q$
to handle $N$ points even if you put one point in a single quadrant
all you need is $log2(N)$planes.

\medskip{}

\textbf{\large OTHER APPLICATIONS }{\large \par}

In all these applications one must some how employ mappings to reduce
multi-dimensional data, temporal data or any other kind of data to
image points in n-dimensional space, it is only then that the methods
of the algorithm described in this paper can be usefully employed
for classification or decision making. An example illustrating this method is given in Ref.[12], by which any $n-$digit prime number can be depicted as a point in $n-$ dimension space; and therefore a prime number repository for storage and easy retrieval of primes can be created. The figure 12, 
 below shows how all 2-digit prime numbers can be considered as points in 2-d space and separated by just 10 planes.

\medskip{}
\begin{figure}[htp]
 \begin{center}
\includegraphics[scale=0.5]{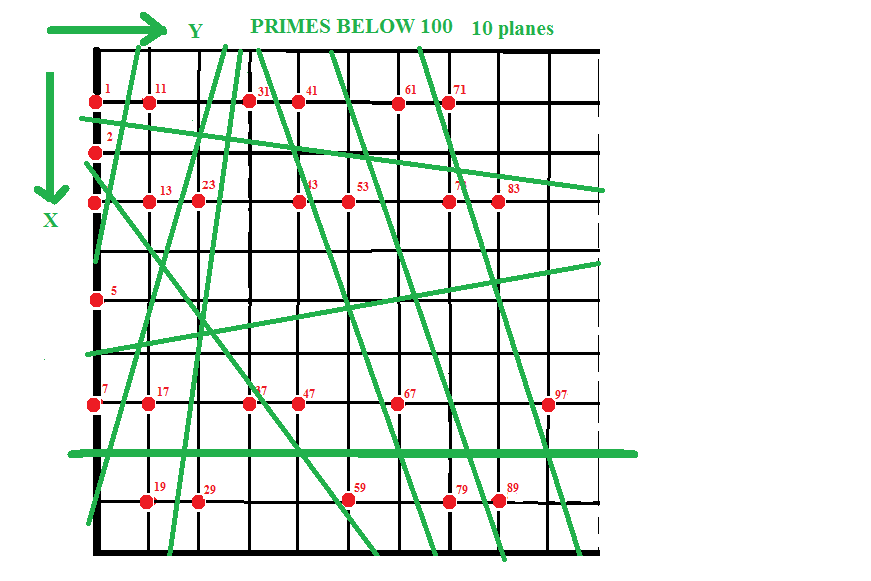}
\caption{Separation of 2-digit primes. Each prime is given a unique (x,y) // coordinate, eg, 53 is put in position (3,5)}

\label{fig:fig-m} 
\end{center}
\end{figure} 
\medskip{}

\textbf{\large 1. Chess Games:}We can treat each chess game as a point
in $2n$-dimension. Its coordinates being $(w1,b1,w2,b2,...,wn,bn)$
the $n$ moves made by white and black (it is not hard to convert
the moves $w1,b1,$ etc. into numbers so the above array which represents
a single chess game, is represented as a single point in 2n-dimension
space). It is possible to store numerous chess games, say $N$, for
easy retrieval using the same technique as described for medical data.
The number of planes involved would be only $q=O(log2N)$. 

\medskip{}

\textbf{\large 2. Storing of sequences:}\textbf{ (for easy retrieval
and for making predictions using historical data)}

In the above cases the data is static and not dynamic. In many life
situations it is necessary to tackle dynamic data viz. sequences,
like a sequence of events, tracking of time signal, or even such matters
as health monitoring of machines, (eg. see Jeff Hawking {[}13{]}-{[}14{]},
who has underlined the importance of storing and recalling a sequence
of events in order to further research in artificial intelligence).

\medskip{}

We then need a method using which we can compare two sequences, say,
one a shorter sequence, $c(1),c(2),c(3),...,c(s)$,
of length, $s$, with a longer sequence $h(1),h(2),h(3),..,h(r)$
of length $r$, which is stored in the memory. The $c(j)$ may be
the `condition\textquoteright{} of a machine in the year, $j$, of
its working life. So that the possible `future` values of the shorter
sequence$c(j)$ can be predicted by comparing it with a longer sequence$h(r)$,
which is the record of some machine which has already lived its life
and whose records are now stored in a memory along with the historical
records of many such machines. We must imagine that the longer second
sequence has been extracted from the data base, by the retrieval engine
(see Sec 3.2), because it happens to have its first $s$ values somewhat
close to the $s$ values of the shorter sequence. 

\medskip{}

We describe a method of mapping a sequence to points in a $n$-dimensional
space. This mapping is necessary to tackle dynamic situations and
tracking/memorizing a sequence of events. The data contained in the
sequence are then `mere\textquoteright{} points in $n$-dimensional
space which are then separated by using the algorithm using $q$ planes,
the Orientation Vectors are used to store the data in a repository
Sec 3.2). But in order to perform all these tasks we need a scheme
to convert a sequence to points in $n$-dimensional space; we now
demonstrate how this can be done. 

We will suppose$n$ is the max number of years that each record has.
It is possible to store the sequence as points in an $n$ dimensional
space using the following scheme. We define the coordinates in $n$-dimension
space for each member of the sequence as follows:

$(c(1),0,0,0,...,0)$ : \qquad{}\qquad{}$n-1$ zeros;

$(c(1),c(2),0,...,0)$:\qquad{}\qquad{}$n-2$ zeros

$(c(1),c(2),c(3),0,0,...,0)$: \qquad{}\qquad{} $n-3$ zeros 

$.........$

$(c(1),c(2),c(3),...,c(s),0,...,0):$ \qquad{}\qquad{}$(n-s)$ zeros 

\medskip{}

Note: Each of the above is a point in $n$-dimension space. Hence
the sequence of points given above is like a world-line in $n$-space.
(Even though, we considered the $c(j)$ as a single number; there
is no difficulty if this is (say) $m$ dimensional \textendash{} then
the actual value of the bigger space will $n.m$ dimensional i.e.
be $n\rightarrow n.m$ . And a similar scheme for the sequence$h$:

$(h(1),0,0,0,...,0)$:\qquad{}\qquad{} $n-1$zeros;

$(h(1),h(2),0,...,0):$\qquad{}\qquad{} $n-2$ zeros

$(h(1),h(2),h(3),0,0,...,0):$\qquad{}\qquad{} $n-3$ zeros

$......$

$(h(1),h(2),c(3),...,h(n):$\qquad{}\qquad{} We assume $r=n$.

\medskip{}

Similarly, we can think of the sequence of points above as a world-line
in n-space. The problem of comparing two sequences $c$ and $h$ has
been reduced to the comparison of two world-lines. Since we have separated
every point such as the $h`s$ by using
planes, we can easily retrieve any sequence such as $h$ by using
the retrieval engine. Now, suppose we are presented a point $P$ (as
described in Sec 3.2), whose coordinates are $(c(1),c(2),c(3),...,c(s),0,...,0)$
; the retrieval engine will retrieve a point $L$ whose coordinates
$(h(1),h(2),h(3),...,h(s),0,...,0),$
are closest to point $P$. This implies you have detected another
world-line $h$, which had had similar experiences in its first$s$
episodes of its `life' as that of world-line $c$; and hence it is
quite possible that the fate of $c$ would be similar to what befell
$h.$ Thus making it possible to retrieve the entire record $(h(1),h(2),c(3),...,h(n)$,
thus enabling us to predict the possible values of $c(j),$ when $j>s$
by looking at the other values $h(s+1),h(s+2),h(s+3),...,h(n)$
which are now been made available. 

\medskip{}

In the above example, you could also think of world line $c$ as the
medical record of some person who is presently living and is $s$
years old. And the world line $h$ as the medical record of some person
who had probably lived and died, but whose first $s$ years of life she/he
had a similar medical history as that of $c$. This is how we can
predict the `future life' of a presently occurring sequence of events
by using the repository containing historical data of sequences that
have occurred in the past.

\medskip{}

\textbf{\large 3. Other Possibilities:} We could similarly convert
decision trees and/or logical trees to sequences which can then be
mapped as points in a large $n$ dimension space, so we see the algorithm
can help in decision making and imitative learning etc. of large complex
data.

\medskip{}

END OF BRIEF ESSAY

\end{document}